\documentclass[a4paper,11pt]{article}
\pdfoutput=1

\usepackage{jcappub}

\usepackage[T1]{fontenc} 

\usepackage{graphicx}
\usepackage{indentfirst}
\usepackage{latexsym}
\usepackage{multirow}
\usepackage{tabls}
\usepackage{epsfig}
\usepackage{multirow}
\usepackage{subfigure}
\usepackage{comment}
\usepackage{hyperref}
\usepackage{color}
\usepackage[usenames,dvipsnames]{xcolor}
\usepackage{amssymb}
\usepackage{amsfonts}
\usepackage{amsmath}
\usepackage{bm}
\usepackage{mathrsfs}
\usepackage{empheq}
\usepackage[normalem]{ulem}

\newcommand{\be}{\begin{equation}}
\newcommand{\ee}{\end{equation}}

\title{A framework for LISA population inference}

\author[a,b]{Alexandre Toubiana}
\author[c]{Jonathan R. Gair}

\affiliation[a]{Dipartimento di Fisica ``G. Occhialini'', Università di Milano-Bicocca, \\Piazza dell'Ateneo Nuovo, 1, 20126, Milano, Italy}
\affiliation[b]{INFN, Sezione di Milano-Bicocca, Piazza della Scienza 3, 20126 Milano, Italy}
\affiliation[c]{Max Planck Institute for Gravitational Physics (Albert Einstein Institute), Am M\"{u}hlenberg 1, 14476, Potsdam, Germany}

\emailAdd{alexandre.toubiana@unimib.it}
\emailAdd{jonathan.gair@aei.mpg.de}


\date{\today}

\abstract{The Laser Interferometer Space Antenna (LISA) is expected to have a source rich data stream containing signals from large numbers of many different types of source. This will include both individually resolvable signals and overlapping stochastic backgrounds, a regime intermediate between current ground-based detectors and pulsar timing arrays. The resolved sources and backgrounds will be fitted together in a high dimensional Global Fit. To extract information about the astrophysical populations to which the sources belong, we need to decode the information in the Global Fit, which requires new methodology that has not been required for the analysis of current gravitational wave detectors. Here, we 
present a hierarchical Bayesian framework to infer the properties of astrophysical populations directly from the output of a LISA Global Fit, consistently accounting for information encoded in both the resolved sources and the unresolved background. Using a simplified model of the Global Fit, we illustrate how the interplay between resolved and unresolved components affects population inference and highlight the impact of data analysis choices, such as the signal-to-noise threshold for resolved sources, on the results. Our approach provides a practical foundation for population inference using LISA data. }

\begin{document}

\maketitle

\section{Introduction}
The hierarchical Bayesian formalism~\cite{Loredo:2004nn,Adams:2012qw,Mandel:2018mve,Vitale:2020aaz} is now routinely applied in gravitational-wave (GW) astronomy to deconvolve measurement errors and selection effects in order to measure the global properties of source populations. It is used extensively in the analysis of LIGO/Virgo/KAGRA (LVK)~\cite{LIGOScientific:2014pky,VIRGO:2014yos,Somiya:2011np,Aso:2013eba} observations to infer the astrophysical properties of stellar-mass compact objects~\cite{Talbot:2018cva,Fishbach:2018edt,LIGOScientific:2025pvj,Antonini:2025ilj,Adamcewicz:2025phm,Tong:2025wpz,Tenorio:2025nyt,Banagiri:2025dmy,Guttman:2025jkv,Ray:2025xti,Tiwari:2025oah,Tong:2025xir,Gennari:2025nho,Heinzel:2024jlc}, to measure cosmological parameters~\cite{Mastrogiovanni:2021wsd,LIGOScientific:2025jau,MaganaHernandez:2025cnu}, and to perform tests of General Relativity~\cite{LIGOScientific:2021sio,Payne:2023kwj,Magee:2023muf,Zhong:2024pwb}. A similar framework is also applied to pulsar-timing-array (PTA)~\cite{Manchester:2010efd,EPTA:2016ndq,2009arXiv0909.1058J} data in order to infer the properties of supermassive black holes in the late Universe~\cite{EPTA:2023xxk,NANOGrav:2023hfp,Toubiana:2024bil}. Future detectors, such as the Laser Interferometer Space Antenna (LISA)~\cite{LISA:2017pwj,LISA:2024hlh}, will require a new population inference paradigm, however. Crucially, while current LVK observations consist exclusively of individually resolved mergers of stellar-mass compact objects, and PTAs observe a stochastic background generated by the superposition of inspiralling supermassive black hole binaries, neither of these regimes will apply to LISA, which will instead observe a superposition of resolved and unresolved signals from the same astrophysical population.

\begin{figure}
    \centering
    \includegraphics[width=1\columnwidth]{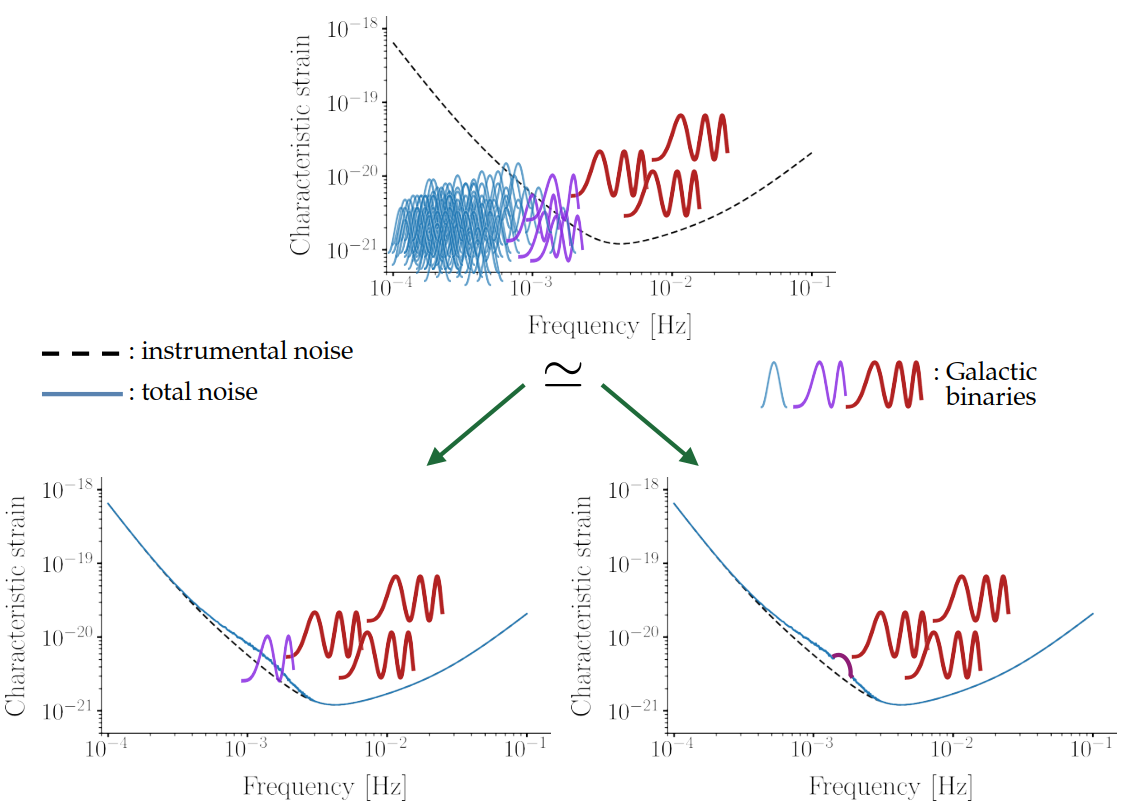}
    \caption{Schematic representation of the Galactic-binary signal in LISA and its decomposition into a stochastic background and a set of resolved sources, shown in terms of the characteristic strain. This signal is continuously present in the LISA datastream. The instrumental noise, shown as dashed lines, corresponds to the SciRDv1 noise curve~\cite{scirdv1}, while the total noise, in blue, includes the Galactic-binary population from~\cite{Toubiana:2024qxc} in the “weak tides, low spins” scenario. Typically, the number density of sources is expected to scale with frequency as $\sim f^{-11/3}$, implying fewer sources at higher frequencies. Moreover, the GW amplitude of an individual source scales as $f^{2/3}$, while the frequency derivative increases as $f^{11/3}$ for GW-dominated evolution, making the signals less monochromatic and easier to distinguish from one another. Altogether, these effects make high-frequency sources more likely to be resolved.  The separation between a stochastic signal and resolved sources is performed within the Global Fit, and different splittings are possible, with sources being effectively “exchanged” between the background and the resolved population. The lower panels illustrate this exchange: the purple source at a few mHz can either be resolved or instead contribute to an increase in the total noise. As we discuss in this work, no data or events are discarded in LISA analyses, and the notion of “selection effects” corresponds instead to determining whether a source is resolved or  allowed to contribute to the stochastic signal.}
    \label{fig:lisa_data}
\end{figure}

Scheduled for launch in 2035, LISA will open a new window onto the gravitational Universe, with a potentially transformative impact on our understanding of the formation and evolution of massive black holes~\cite{Gair:2010bx,Sesana:2010wy,Klein:2015hvg,Dayal:2018gwg,Barausse:2020mdt,Toubiana:2021iuw,Chen:2022sae,Fang:2022cso,Spadaro:2024tve,Langen:2024ygz,Toubiana:2024bil}, the population of Galactic binaries (GBs)~\cite{Nelemans:2001hp,Ruiter:2007xx,Korol:2017qcx,Lamberts:2019nyk,Korol:2021pun,Biscoveanu:2022sul,Toubiana:2024qxc,Delfavero:2024zyl}, the abundance and properties of extreme-mass-ratio inspirals (EMRIs)~\cite{Gair:2010yu,Amaro-Seoane:2012lgq,Babak:2017tow,Pan:2021ksp,Seoane:2024nus}, and the environments of stellar-mass black hole binaries~\cite{Sesana:2016ljz,Toubiana:2020drf,Sberna:2022qbn,Zwick:2025wkt,Zwick:2025qzv,Chen:2025qyj}.
LISA is expected to observe tens of massive black hole mergers, hundreds of EMRIs and $\sim 10^7$ GBs. These sources are all long-lived and nearly monochromatic, so the GW signals in the LISA data will be overlapping almost entirely in time and frequency~\cite{Nelemans:2001hp,Ruiter:2007xx,Korol:2017qcx,Lamberts:2019nyk,Korol:2021pun,Toubiana:2024qxc,Delfavero:2024zyl}, as illustrated schematically in Fig.~\ref{fig:lisa_data}. It will be necessary to perform a Global Fit in which the number and properties of all sources of all types, as well as those of the noise, are inferred simultaneously~\cite{Cornish:2005qw,Littenberg:2023xpl,Katz:2024oqg,Strub:2024kbe,Deng:2025wgk}.

The true model of the data is a superposition of individual binaries. At low frequencies, the number of overlapping signals is so large that sources can barely be distinguished. Attempting to resolve individual signals would lead to a highly degenerate posterior distribution that would be nearly impossible to sample. However, by virtue of the central limit theorem, in this regime the signal is well described as a Gaussian stochastic process, with phases varying randomly from bin to bin. At higher frequencies, sources are rarer and less monochromatic, and this approximation breaks down: the signal is neither Gaussian nor uncorrelated between neighbouring bins, as these can be dominated by the same individual binary. The transition between the two regimes is in fact smooth, and in this intermediate region the data are better described by more complex, non-Gaussian stochastic background models~\cite{Karnesis:2024pxh}. Thus, the GB signal is traditionally divided into two components: (i) a stochastic background, not necessarily Gaussian, which dominates over the instrumental noise near $1 \ \mathrm{mHz}$, and (ii) a set of resolved sources, with an expected number of $\sim 10^4$, which can be individually characterised. This separation into two components is illustrated in Fig.~\ref{fig:lisa_data}, where the contribution from the majority of GBs is absorbed into the total noise curve, while a smaller number of louder, higher-frequency, and less monochromatic binaries are individually resolved.

These two components are fitted jointly in the Global Fit, leading to correlations between the inferred properties of the resolved population and the estimates of the total noise level. A similar mixed scenario may arise for EMRIs as well. Their intrinsic rate remains uncertain enough that LISA may detect anything from a handful of events to a population large enough to contribute significantly to the total stochastic background~\cite{Gair:2010yu,Amaro-Seoane:2012lgq,Babak:2017tow,Pan:2021ksp,Seoane:2024nus,Bonetti:2020jku,Pozzoli:2023kxy}.
Thus, a key challenge is to understand how the output of the Global Fit can be translated into constraints on the underlying astrophysical population of LISA sources, a question that remains open. It is important to note that the split into resolved events plus background is a choice of the (approximate) model used to fit the data. When we make analysis choices we change which events are resolved and which are lumped into the background, but these models are only approximate and, in practice, the Global Fit explores different configurations, with some sources sometimes being absorbed into the background and sometimes being resolved, particularly those lying near the region of parameter space where the transition between unresolved and resolved occurs, as illustrated in Fig.~\ref{fig:lisa_data}. Setting the threshold for resolving sources too high means that what is left in the background is not described by a simple stochastic background model. Similarly, setting the threshold too low means many parameters have to be sampled, so there is a trade off between computational cost and the faithfulness of the representation of the data.

The hierarchical Bayesian formulations for population inference usually treat either (i) a set of statistically independent, individually resolved events, or (ii) a stochastic signal that can be related to the underlying source population. However, they do not accommodate situations where a single population contributes simultaneously to a resolvable population and to an unresolved confusion background - precisely the case for LISA GBs, and potentially for EMRIs.
A formalism that includes both resolved events and stochastic backgrounds was proposed in the LVK context~\cite{Callister:2020arv}, but it does not treat resolved and unresolved sources in a fully consistent way. As a consequence, it does not account for the correlations that naturally emerge in a LISA-like Global Fit, particularly the fact that the number of resolvable events is not fixed and that resolving more or fewer sources directly impacts the inferred background. As LISA data-analysis tools are being developed, it is becoming essential to establish a unified formalism capable of treating such populations, both to guide methodological development and to define the data products that should be provided by the collaboration. This issue is also relevant for other space-based interferometers, such as the planned mission TianQin~\cite{TianQin:2020hid}, as well as Taiji~\cite{Ruan:2018tsw} and DECIGO~\cite{Kawamura:2006up}, which are currently under development. It is likewise important for third-generation ground-based detectors, such as the Einstein Telescope~\cite{Punturo:2010zz} and Cosmic Explorer~\cite{Reitze:2019iox}, expected to come online in the next decade, where overlapping signals are anticipated to produce some level of source confusion~\cite{ET:2025xjr,Bavera:2021wmw,Evans:2021gyd,Zhou:2022nmt,Zhong:2022ylh,Zhong:2024dss,Johnson:2024foj}. Finally, similar challenges arise for pulsar timing arrays, which are expected to begin resolving individual sources in the near future~\cite{Sesana:2008xk,Sesana:2010mx,Kelley:2017vox,2017NatAs...1..886M}, particularly with the advent of the Square Kilometre Array~\cite{Janssen:2014dka}.

In this paper, we develop a more general hierarchical Bayesian framework capable of processing the output of a Global Fit in order to describe populations that generate both a stochastic background and individually resolvable signals. In short, the problem can be thought of as involving two populations: one that contributes to the background and another that gives rise to resolvable sources. Crucially, these populations overlap in parameter space, and we derive how to account for this in a self-consistent manner when evaluating the population likelihood in Sec.~\ref{sec:hba_lisa}, after reviewing the approaches that are currently employed for LVK and PTA analyses in Sec.~\ref{sec:std_formalism}. In the appendix, we discuss the conditions under which our formulation reduces to the standard hierarchical Bayesian approach and how it differs from that of~\cite{Callister:2020arv}. Finally, using a toy model Global Fit, we illustrate how the framework can be applied in practice and highlight the interplay between the stochastic and resolved components in constraining the underlying astrophysical population. We stress that the toy model and its implications for LISA data analysis are very understandable even without the derivations presented in Sec.~\ref{sec:hba_lisa}. We therefore encourage readers interested in LISA, but less in the statistical framework for population inference, to read this section independently of Sec.~\ref{sec:hba_lisa}.

\section{Current hierarchical Bayesian formalism}\label{sec:std_formalism}
We start by recalling the hierarchical Bayesian formalism used to analyse LVK observations.  We refer to Refs.~\cite{Loredo:2004nn,Mandel:2018mve,Vitale:2020aaz}
for more details, and review here the main steps and results.

We assume that data $d$ is  generated by events with
parameters $\theta$, and that these events are drawn from a population
with differential number density $N(\theta|\Lambda)$, characterised by a set of hyperparameters $\Lambda$ we wish
to infer.
The total expected number of events is
\begin{equation}
    N(\Lambda) = \int N(\theta|\Lambda)\,{\rm d}\theta \, .
\end{equation}
For the sake of compactness we will now drop the $\Lambda$ dependence in $N$. We also introduce the population prior on the event parameters:
\begin{equation}
    p(\theta|\Lambda) = \frac{N(\theta|\Lambda)}{N(\Lambda)} \, .
\end{equation}

The full datastream $d$ is divided into chunks, and only those chunks
for which a detection statistic passes a selection criterion, e.g.,~a false-alarm rate below 1 or a signal-to-noise ratio above 8, are used for the analysis.
We assume that at most one astrophysical event contributes to each data chunk and that, over the observing time, $n_{\rm det}$ events were detected, and an additional
$n_{\rm ndet}$ astrophysical events occurred but were not detected, because the data in the chunks in which they occurred did not pass the selection criterion.

Let $\{d_c\}_{n_{\rm det}}$ denote the
signal-containing data chunks for which the detection statistic passed the selection criterion and $\{\theta\}_{n_{\rm det}}$ denote the corresponding parameters of the events in those data chunks, and let $\{d_c\}_{n_{\rm ndet}}$ and
$\{\theta\}_{n_{\rm ndet}}$ denote the corresponding quantities for the data chunks containing signals which did not pass the selection criterion. The joint probability for all (signal-containing) data chunks and corresponding event parameters takes the form
\begin{equation}
    p(\{d_c\}_{n_{\rm det}},\{\theta\}_{n_{\rm det}},\{d_c\}_{n_{\rm ndet}},\{\theta\}_{n_{\rm ndet}}|\Lambda) \propto e^{-N}\prod_{i=1}^{n_{\rm det}} p(d_{c,i}|\theta_i)N(\theta_i|\Lambda) \prod_{j=1}^{n_{\rm ndet}} p(d_{c,j}|\theta_j)N(\theta_j|\Lambda). \label{eq:full_std}
\end{equation}
Here and in the following the proportionality symbol accounts for data-dependent normalisation factors that we can discard.
Marginalising over the single event parameters and over the datasets with undetected events, we get
\begin{equation}
    p(\{d_c\}_{n_{\rm det}}|\Lambda) \propto e^{-\xi(\Lambda)N}\prod_{i=1}^{n_{\rm det}} \int p(d_{c,i}|\theta) N(\theta_i|\Lambda) \
     {\rm d} \theta  .\label{eq:pop_std}
\end{equation}
where we have introduced the selection function
\begin{align}\label{eq:pdet}
    \xi(\Lambda)=\int  p_{\rm det}(\theta)p(\theta|\Lambda)\  {\rm d} \theta,\\
    p_{\rm det}(\theta)=\int_{d_c \in {\rm det}} p(d_c|\theta) \ {\rm d } d_c.
\end{align}
The integral over $d_c$ is performed over data chunks passing the selection criterion.

We now discuss the case where the observable is a stochastic signal. Let $\Sigma$ denote the parameters of that signal.
The likelihood for observing the datastream $d$ conditioned on the population parameters is
\begin{equation}\label{eq:pop_stochatic}
    p(d|\Lambda)
    = \int p(d|\Sigma)\, p(\Sigma|\Lambda)\, {\rm d}\Sigma \, .
\end{equation}

If we assume that the measurement of individual events and of the stochastic noise are independent, a joint likelihood using both pieces of information can be obtained by multiplying Eq.~\eqref{eq:pop_std} and Eq.~\eqref{eq:pop_stochatic}, see Ref.~\cite{Callister:2020arv}:
\begin{equation}\label{eq:pop_res_stoch_std}
    p(d|\Lambda)
    \propto e^{-\xi(\Lambda)N}
      \prod_{i=1}^{n_{\rm det}}
      \int p(d_{c,i}|\theta,{\rm obs})\, N(\theta_i|\Lambda)\,
      {\rm d}\theta
      \times
      \int p(d|\Sigma)\, p(\Sigma|\Lambda)\, {\rm d}\Sigma \, .
\end{equation}
In the above, the second term dependence on the full data is only an approximation, as it includes also all of the resolved-source containing segments, $\{ d_c \}_{n_{\rm det}}$. This is acceptable for current ground-based detectors because data chunks containing detectable events are much rarer than those without. We can therefore assume that these rare events do not affect the reconstruction of the stochastic noise, and conversely, that the noise can be considered known when these events are observed, and does not need to be constrained jointly. However, as we will show in App.~\ref{app:lvk_limit}, this expression does not fully incorporate the information of resolved and unresolved sources self-consistently . It is a good approximation for the current situation, where we can only place upper bounds on the background. We also refer to Sec.~IV of the Supplemental Material of Ref.~\cite{Zhong:2025qno} for a discussion on the validity of this approximation for third-generation ground-based detectors.

In the LISA data these approximations do not hold. Firstly, there will be no selection of data chunks that are interesting, but all observed data will be included and analysed. In addition, sources will be in band for a long time (continuously for GBs) and the stochastic signal they form, in addition to the instrumental noise, must be inferred jointly with the number and properties of the resolvable sources. This can be achieved by means of reversible-jump Markov chain Monte Carlo (RJMCMC)~\cite{10.1093/biomet/82.4.711}, where the sampler adds and removes sources while simultaneously sampling over the model parameters. Different RJMCMC samples therefore have different dimensionality. Samples including more resolved sources will have fewer sources contributing to the background, and vice versa, so that the number and properties of resolved sources will be correlated with the properties of the corresponding background, necessitating that the samples of both the resolvable and stochastic components be treated jointly in the hierarchical inference. We will now discuss how this can be addressed.

\section{General hierarchical Bayesian formalism}\label{sec:hba_lisa}

We seek to model jointly the contribution of individual sources to the resolvable sources and to the stochastic noise, while accounting for the fact that the number of sources, resolved and unresolved, is unknown. We start by writing the joint probability of data, individual events, number of sources, and noise parameters:
\begin{equation}
    p(d,\{\theta\}_n,n,\Sigma|\Lambda) \propto p(d|\{\theta\}_n,n,\Sigma)p(\Sigma|\{\theta\}_n,n,\Lambda)\prod_{i=1}^n N(\theta_i|\Lambda)e^{-N}. \label{eq:gen_form}
\end{equation}
The above equation states that the likelihood for the data $d$ depends on the events and on the noise properties, and that the noise properties can also depend on the events drawn from $N(\theta|\Lambda)$. We will ultimately marginalise over $\{\theta\}_n$, $n$ and $\Sigma$ to obtain the population likelihood.
The noise parameters $\Sigma$ can also contain information about quantities that do not depend on the source population, e.g., instrumental noise, but we keep the same notation for generality. For this same reason, the probability of $\Sigma$ is also conditioned on $\Lambda$, allowing us to capture both the components of the noise that depend on the population through $(\{\theta\}_n,n)$ and those that are independent of the astrophysical model being considered. Note that throughout the derivation, we make no assumption about the likelihood, $ p(d|\{\theta\}_n,n,\Sigma)$, being Gaussian.

Next, we want to account for the fact that, when analysing the datastream, not all individual sources $\{\theta\}_n$ enter the likelihood in the same way. In principle, one could attempt to recover all GBs individually, as this is a true description of the data. However, at low frequencies, the very small frequency derivatives lead to a strong confusion between low signal-to-noise ratio (SNR) sources, making the problem highly degenerate and causing any sampler to struggle to correctly represent the posterior. To alleviate this problem, we do not try to resolve the large number of low SNR sources, but instead represent them as a stochastic background. Often the background is assumed to be Gaussian, but the formalism we derive does not rely on this assumption. How we split the sources between resolved and stochastic components is somewhat arbitrary, and represents a choice of a specific approximation to the true likelihood, and depends on what priors we place on the parameters of the resolvable GBs, as we will explore in Sec.~\ref{sec:toy_model}.

Let us now describe how to incorporate this model into the hierarchical Bayesian formalism. The derivation below leads to the main result of this paper, Eq.~\eqref{eq:hyper_logl_1}.

\subsection{Building intuition from a simpler setting: the separable case}\label{sec:sep_case}

Let us first assume that there is no ambiguity whether a source is resolvable or not based on its properties, e.g,.~its frequency and/or its amplitude. This is not what happens in practice, but it will be easier to start with this case and afterwards consider how things change in the more realistic scenario.

The perfect separability assumption means that the astrophysical population is partitioned into a population of resolvable sources with differential rate $N_1(\theta|\Lambda)$ and a population of non-resolvable sources with differential rate $N_2(\theta|\Lambda)$. We denote the support of these populations on parameter space as $\mathcal{S}_{1,2}$, assuming that they have no intersection and form a complete partition of the parameter space. From standard properties of the Poisson distribution, thie two populations can be modelled as independent Poisson processes.

 For a given $\{\theta\}_{n}$ set there are $n_{1}(\{\theta\}_{n},\Lambda)$ events in the support of $\mathcal{S}_{1}$ and $n_2=n-n_1$ in $\mathcal{S}_{2}$. Marginalising over $\{\theta\}_n$ then amounts to marginalising over $n_{1}(\{\theta\}_{n},\Lambda)$ and over the parameters of these sources, $\{\theta\}^1_{n_1}$, as well as over those in $\mathcal{S}_{2}$, $\{\theta\}^2_{n_2}$. Then, marginalising over $n$ in turn, amounts to marginalising over $n_2$. Finally, marginalising over $\Sigma$ as well, we get
\begin{align}
    p(d|\Lambda) \propto & \sum_{n_1=0}^{+\infty} \sum_{n_2=0}^{+\infty} \frac{N^{n_1}N^{n_2}}{n_1!n_2!}e^{-N} \int {\rm d} \{\theta\}^1_{n_1} {\rm d}\{\theta\}^2_{n_2} {\rm d}\Sigma \  p(d|\{\theta\}^1_{n_1},n_1,\Sigma) p(\Sigma|\{\theta\}^2_{n_2},n_2,\Lambda)  \nonumber \\ & \times \prod_{i=1}^{n_1}p(\theta^1_i|\Lambda)\prod_{j=1}^{n_2}p(\theta^2_j|\Lambda)  . \label{eq:hyper_logl_0}
\end{align}
We split the product over $n$ sources as a product over $n_1$ resolvable and $n_2$ unresolvable sources, and divide by the corresponding factorial terms to avoid overcounting when marginalising. This is because, in our approach, resolvable and unresolvable sources are distinguishable from each other, but sources within each class are indistinguishable once we marginalise.
Defining
\begin{equation}\label{eq:probs_sep}
    p_{1,2}(\theta|\Lambda)=\frac{p(\theta|\Lambda)\Xi_{1,2}(\theta)}{\int_{\mathcal{S}_{1,2}}p(\theta|\Lambda)\ {\rm d}\theta},
\end{equation}
where $\Xi_{1}(\theta)$ is 1 if $\theta \in \mathcal{S}_1$ and 0 otherwise, and reciprocally for $\Xi_2$, we have
\begin{align}
   & N_{1,2}(\theta|\Lambda)=N_{1,2}p_{1,2}(\theta|\Lambda) \\
    &N_{1,2}= \int_{\mathcal{S}_{1,2}} N(\theta|\Lambda).
\end{align}
We can then write Eq.~\eqref{eq:hyper_logl_0} as
\begin{align}
    p(d|\Lambda)  \propto & \sum_{n_1=0}^{+\infty} \sum_{n_2=0}^{+\infty} \frac{N_1^{n_1}N_2^{n_2}}{n_1!n_2!}e^{-N_1-N_2}  \int {\rm d} \{\theta\}^1_{n_1} {\rm d}\{\theta\}^2_{n_2} {\rm d}\Sigma \  p(d|\{\theta\}^1_{n_1},n_1,\Sigma) p(\Sigma|\{\theta\}^2_{n_2},n_2,\Lambda) \nonumber \\
    & \times \prod_{i=1}^{n_1}p_1(\theta^1_i|\Lambda)\prod_{j=1}^{n_2}p_2(\theta^2_j|\Lambda) , \label{eq:hyperl_1}
\end{align}
where we divided the $p(\theta^{1,2})$ terms by the normalisation of Eq.~\eqref{eq:probs_sep} to obtain $p_{1,2}(\theta^{1,2})$, and absorbed the corresponding normalisation factors into the $N^{n_{1,2}}$ terms.

\subsection{General case}
Now, we consider the more general case where a source may sometimes be resolved and sometimes not. We introduce a function $\alpha(\theta,\Lambda)$ to represent the probability that a source with parameters $\theta$ is resolved, i.e., that it belongs to population 1. We then have
\begin{align}
& N_1(\theta|\Lambda)=\alpha(\theta,\Lambda)N(\theta|\Lambda)  \\
& N_2(\theta|\Lambda)=(1-\alpha(\theta,\Lambda))N(\theta|\Lambda)   \\
& N_{1,2}=\int N_{1,2}(\theta|\Lambda) \ {\rm d} \theta, \\
\end{align}
The function $\alpha(\theta,\Lambda)$ is imposed by the analyst, not by the Universe, and determines the approximate model used to describe the data. It should be chosen such that the induced model of background + resolvable sources is a good description of the data. The right choice of $\alpha(\theta,\Lambda)$ might depend on the population, so we include $\Lambda$ in $\alpha$. The case considered in Sec.~\ref{sec:sep_case} corresponds to the choice $\alpha(\theta,\Lambda)=\Xi_1(\theta)$.
Going back to Eq.~\eqref{eq:gen_form}, for a given set $\{\theta\}_n$, there are now many possible ways of splitting the sources between the two populations. A particular split $(\{\theta\}^1_{n_1},\{\theta\}^2_{n-n_1})$ has probability
\begin{equation}\nonumber
\prod_{i=1}^{n_1} \alpha(\theta^1_i,\Lambda) \prod_{j=1}^{n-n_1}(1-\alpha(\theta^2_j,\Lambda)).
\end{equation}

When marginalising over $\{\theta\}n$, we obtain an expression analogous to Eq.~\eqref{eq:hyper_logl_0}, but where the integrand is now weighted by the split probabilities:
\begin{align}
&p(d|\Lambda)= \sum_{n_1=0}^{+\infty} \sum_{n_2=0}^{+\infty} \frac{N^{n_1}N^{n_2}}{n_1!n_2!}e^{-N}
\int {\rm d} {\theta}^1_{n_1} {\rm d}{\theta}^2_{n_2} {\rm d}\Sigma \
p(d|\{\theta\}^1_{n_1},n_1,\Sigma)
p(\Sigma|\{\theta\}^2_{n_2},n_2,\Lambda) \nonumber \\
& \times \prod_{i=1}^{n_1}p(\theta^1_i|\Lambda)\alpha(\theta^1_i,\Lambda)
\prod_{j=1}^{n_2}p(\theta^2_j|\Lambda)(1-\alpha(\theta^2_j,\Lambda)) .
\label{eq:hyper_logl_2}
\end{align}
Defining
\begin{align}
& p_{1} (\theta|\Lambda)= \frac{p(\theta|\Lambda)\alpha(\theta,\Lambda)}{\int p(\theta|\Lambda) \alpha(\theta,\Lambda) \ {\rm d} \theta} \label{eq:p1def}\\
& p_{2} (\theta|\Lambda)=\frac{p(\theta|\Lambda)(1-\alpha(\theta,\Lambda))}{\int p(\theta|\Lambda) (1-\alpha(\theta,\Lambda)) \ {\rm d} \theta},
\label{eq:p2def}
\end{align}
we recover Eq.~\eqref{eq:hyperl_1}.
Finally, we define
\begin{align}
    & p(\Sigma|n_2,\Lambda)=\int {\rm d} \{\theta\}^2_{n_2} \ p(\Sigma|\{\theta\}^2_{n_2},n_2,\Lambda)\prod_{j=1}^{n_2}p_2(\theta^2_j|\Lambda), \\
    & p(\Sigma|\Lambda)=\sum_{n_2=0}^{+\infty} p(\Sigma|n_2,\Lambda) \frac{N_2^{n_2}}{n_2!}e^{-n_2},  \label{eq:hyperlogl_marg}
\end{align}
and get
\begin{empheq}[box=\fbox]{equation}
    p(d|\Lambda) \propto \sum_{n_1=0}^{+\infty} \int {\rm d} \{\theta\}^1_{n_1} {\rm d}\Sigma \  p(d|\{\theta\}^1_{n_1},n_1,\Sigma) p(\Sigma|\Lambda)\prod_{i=1}^{n_1}p_1(\theta^1_i|\Lambda) \frac{N_1^{n_1}}{n_1!}e^{-N_1} \label{eq:hyper_logl_1}.
\end{empheq}
This is main result of this paper, the general equation to be used for hierarchical Bayesian inference for LISA data. We will illustrate this with a single population chosen to represent GBs, but this can be readily extended to multiple populations, for example massive black hole binaries and EMRIs. We stress that our formalism does not rely on the likelihood used in the Global Fit ($p(d|\{\theta\}^1_{n_1},n_1,\Sigma)$) being Gaussian, and heavier-tailed likelihoods could be used~\cite{Sasli:2023mxr,Karnesis:2024pxh}. In practice, when running a Global Fit with RJMCMC, we have samples from $p(\{\theta\}^1_{n_1},n_1,\Sigma|d)$. Using Bayes' theorem, and denoting by $\pi_{\rm PE}(\{\theta\}^1_{n_1},n_1,\Sigma)$ the prior used to obtain the posterior during initial sampling, we have
\begin{equation}
    p(d|\Lambda) \propto\sum_{n_1=0}^{+\infty} \int {\rm d} \{\theta\}^1_{n_1} {\rm d}\Sigma \ \frac{ p(\{\theta\}^1_{n_1},n_1,\Sigma|d)p(d)}{\pi_{\rm PE}(\{\theta\}^1_{n_1},n_1,\Sigma)} p(\Sigma|\Lambda)\prod_{i=1}^{n_1}p_1(\theta^1_i|\Lambda) \frac{N_1^{n_1}}{n_1!}e^{-N_1} \label{eq:marg_hyperl}.
\end{equation}
This can be evaluated via Monte-Carlo integration as
\begin{equation}
p(d|\Lambda)\propto \frac{1}{N_s} \sum_{k=1}^{N_s} \frac{\prod_i^{n_{1,k}}p_1(\theta^1_{i,k}|\Lambda)p(\Sigma_k|\Lambda)}{\pi_{\rm PE}(\{\theta\}^1_{n_1,k},\Sigma_k)}  \frac{N_1^{n_{1,k}}}{n_{1,k}!}e^{-N_1} ,\label{eq:lisa_pop_rjmcmc}
\end{equation}
where the sum runs over $N_s$ samples of $(\{\theta\}_{n_1},n_1,\Sigma)$ obtained with RJMCMC. We note that here we are implicitly assuming that the prior on individual event parameters used for the initial parameter estimation was fixed. An alternative strategy would be to do the initial parameter estimation including a population model, i.e., introducing hyperparameters into the individual source priors and simultaneously sampling on both the hyperparameters and the event parameters, as recently suggested in the LVK context~\cite{Mancarella:2025uat}.

The resolvability function $\alpha(\theta,\Lambda)$ is not the same as the selection function defined in Eq.~\eqref{eq:pdet}. As discussed in Sec.~\ref{sec:std_formalism}, the selection function arises from marginalisation over undetected events. By contrast,
here we are not discarding events: as we mentioned before, we are choosing how each event contributes to the likelihood, in particular depending on the minimum SNR that we require for sources to be considered resolved. For example, if one chooses a split that effectively forbids resolved sources by raising the SNR threshold, a larger fraction of the data must be described as part of the background. However, at high frequencies this signal is not well described by a Gaussian background, and a likelihood that allows for a non-Gaussian background would then be required (see, e.g., \cite{Karnesis:2024pxh}); otherwise, the inference will be biased. Jointly fitting the background and resolved sources provides an effective way to improve the description of the data and to capture the more “popcorn-like” behaviour of the signal at high frequencies. Crucially, the resolvability function $\alpha(\theta,\Lambda)$ need not be pre-decided. One option is to choose a parametric form with free parameters that enter the hyperparameters $\Lambda$ inferred at the hierarchical level. We now consider a toy-model Global Fit to demonstrate the application of our method and to explore the impact of the choice of the resolvability criterion on the population inference.

\section{Application to a toy model Global Fit}\label{sec:toy_model}

We now consider an example to illustrate the formalism presented here. Our toy model Global Fit is inspired by the expected properties of GBs, but we do not seek to describe an astrophysically realistic population and our mock datasets are much simpler than the LISA one, as they are meant for illustrative purposes only. We construct a population that consists of many individual events, but with a number density that is frequency dependent.

\subsection{Mock Global Fit}
We work in the frequency domain and consider $N_{\rm bin}$ frequency bins equally spaced between $f_{\rm min}$ and $f_{\rm max}$. We denote the bin width by \mbox{$\delta f = (f_{\rm max}-f_{\rm min})/N_{\rm bin}$}. We assume that our sources are monochromatic, i.e., each contributes to a single bin, and have GW strain
\begin{equation}
    h(f) = A_0 \left(\frac{f}{f_{\rm min}}\right)^{2/3} e^{i\varphi} \label{eq:def_strain}
\end{equation}
where $A_0$ is a global constant. The $f^{2/3}$ scaling is the leading-order term in the post-Newtonian expansion, corresponding to the quadrupole GW strain from a circular Keplerian binary. This is expected to be dominant contribution to the strain for Galactic binaries.
The total strain in each bin is the sum of the GW strains of the sources that lie in that bin:
\begin{equation}
    H_i = \sum_{j=1}^{n} h_j \, I(b(f_j)-i) , \label{eq:sum_hs}
\end{equation}
where $n$ is the total number of sources, $b(f)={\rm floor}([f-f_{\rm min}]/\delta f)+1$ is a function indicating to which bin a source belongs,
and $I(b(f_j)-i)$ is an indicator function that is $1$ if $b(f_j) = i$ and $0$ otherwise.
We distribute the sources across frequencies following a power-law with spectral index $\gamma$. We denote by $N_{\rm av}$ the mean number of sources, and draw the number of sources in our data set from a Poisson distribution with this rate.
The differential number of events is assumed to follow a powerlaw
\begin{equation}
   N(f|n,\gamma)=\frac{n(\gamma+1)}{f_{\rm max}^{\gamma+1}-f_{\rm min}^{\gamma+1}} f^{\gamma}
\end{equation}
If the number of sources in each bin is ``large enough'', they form a confusion noise. The variance of the amplitude of the strain in a given bin is
\begin{align}
\sigma^2_{B,0}(f) &= \frac{n_2(\gamma+1)}{f_{\rm max}^{\gamma+1}-f_{\rm min}^{\gamma+1}} A_0^2 f^{4/3+\gamma}\delta_f f_{\rm min}^{-4/3}, \label{eq:sigma_foreground_th}
\end{align}
where $f_i$ is the central bin frequency and B stands for background. We add instrumental noise into each frequency bin, modelling it as white Gaussian noise with a standard deviation $\sigma_n$.

The full procedure used to generate data is:
\begin{enumerate}
\item Draw the number of astrophysical sources $n$ from a Poisson distribution with mean $N_{\rm av}$.
\item Draw $n$ values of $f$ from a power-law distribution with spectral index $\gamma$.
\item Draw $n$ phases randomly in $[-\pi,\pi]$.
\item Compute the total strain in each bin using Eq.~\ref{eq:sum_hs}.
\item Draw $N_{\rm bin}$ points from a Gaussian distribution with standard deviation $\sigma_n$ and $N_{\rm bin}$ phases randomly in $[-\pi,\pi]$, and add the resulting complex numbers to the total strain in each bin.
\end{enumerate}
For the simulated data, we use the following parameters:
\begin{itemize}
    \item $\sigma_n=1$,
    \item $A_0= 3$,
    \item $\gamma = -4$,
    \item $N_{\rm av}=10^6$,
    \item $f_{\rm min}=10^{-4} \ {\rm Hz}$,
    \item $f_{\rm max}=10^{-2} \ {\rm Hz}$.
\end{itemize}
We choose a value of $\gamma=-4$ in order to obtain a reasonable number of resolved sources: enough for the resolved population to provide meaningful information, but not so many as to make the problem computationally prohibitive. At the same time, this choice leaves a sufficiently large portion of the frequency band that can be described by a stochastic background, while remaining close to the expected physical value of $-11/3$ for a population evolving purely through GW emission.

We perform a mock Global Fit using the RJMCMC implementation of the \texttt{Eryn} sampler\footnote{Publicly available at \url{https://github.com/mikekatz04/Eryn}.}~\cite{Karnesis:2023ras}. We fit the data using a Gaussian likelihood
\begin{equation}
   p(d|\{\theta\},\Sigma)= \prod_i^{n_{\rm bin}} \frac{1}{\sigma^2_{B}(f_i)+\sigma_n^2}\exp \left [-\frac{|d_i-H_i|^2}{\sigma^2_{B}(f_i)+\sigma_n^2} \right].
\end{equation}
At high frequency, as the number of sources decreases, we no longer expect the data to be well represented by a Gaussian noise. The background should then vanish, and sources will more likely be resolvable. This is related to the probability of resolving individual sources introduced above, $\alpha(f,\Lambda)$. We use the phenomenological form
\begin{equation}\label{eq:res_func}
\alpha(f,f_{{\rm res}},s_{{\rm res}})= \frac{1}{2} \left [ 1+\tanh\left ( \frac{\log(f)-\log(f_{{\rm res}})}{s_{{\rm res}}} \right ) \right ].
\end{equation}
This induces the background model
\begin{align}
\sigma^2_{B}(f)=A_B^2 f^{4/3+\gamma_B}(1-\alpha(f,f_{{\rm res},B},s_{{\rm res},B})).
\end{align}
For simplicity, we assume the instrumental noise level, $\sigma_n$, to be known. The noise parameters are $\Sigma=(A_B,\gamma_B,f_{{\rm res},B},s_{{\rm res},B})$, for which we adopt flat priors.

For the sampling over resolvable sources, we use $\theta_{\rm PE}=(f,\varphi,\rho)$, where $\rho$ is the SNR of the source. The amplitude of a single source is
\begin{equation}
A=\rho \sqrt{\sigma^2_{B}(f)+\sigma_n^2}.
\end{equation}
Thus, in sampling we do not assume that we know the amplitudes of individual sources scale as
$A_0 f^{2/3}$, with $A_0$ being common to all sources. We found that parameterising in terms of $\rho$ helps avoid proposing samples with amplitudes that are either too small or too large. In practice, this induces a frequency-dependent prior on the source amplitude, leading to proposals that are better adapted to the data. We use flat priors on the parameters of resolved sources. The maximum SNR is set to $\rho_{\rm max}=100$ and we explore different choices for the lower bound on SNR, $\rho_{\rm min}$, to assess its impact on the reconstruction of the mock Global Fit and subsequent population inference. The prior on the number of resolved sources is flat between 0 and 20. In order to avoid degenerate solutions in which multiple sources are added in the same bin and interfere destructively, we limit the number of resolved sources per bin to one. Note that in the formalism as we presented it, it was the function $\alpha(f,\Lambda)$ that controlled the separation between resolved and unresolved sources. In this example we control this via the choice of $\rho_{\rm min}$, but allow $\alpha(f,\Lambda)$ to vary flexibly to account for this. We expect to find that $\alpha(f,\Lambda)$ transitions from $0$ to $1$ around the point where the SNR reaches $\rho_{\rm min}$. Limiting to one source per bin could reduce the ability of the resolved source model to absorb non-Gaussian features of the resolved population if the prior on the amplitude of the sources was not wide enough. Since we use $\rho$ and not $A$ as sampling parameter, effectively allowing for very large amplitudes where the background is higher, this is not a problem here.

\subsection{Population analysis}
We now describe the model used to perform the population analysis. Within our simple model, our expectation is that the information from the resolved sources will help to measure the amplitude distribution of individual sources, and to disentangle the contributions from the number of events and from the single-event amplitudes in the measurement of the background amplitude. The slope of the background distribution, $\gamma_B$, is a population parameter. The parameters $(f_{{\rm res},B},s_{{\rm res},B})$ are really sampling choices, but as we are fitting them to the data they do also tell us properties of the population, namely when the population starts to be well described by a Gaussian. These parameters are common between the background and resolved source population. We should therefore set them to be equal. In practice, however, we found that the sampling at the population analysis stage was more efficient if we allowed these common parameters to be different between the two populations. We model the background parameters as Gaussian distributed with means equal to the parameters characterising the resolved population, $(\gamma_P,f_{{\rm res},P},s_{{\rm res},P})$, and with standard deviations $(\sigma_\gamma,\sigma_f,\sigma_s)$.
The latter quantify the degree to which the information on the population parameters inferred from the resolvable population and from the background disagree under our population model. They are expected to be zero if there is no such discrepancy.

The total number of sources, $n$, is assumed to follow a Poisson distribution with mean $N$. We recall that $n_1$ and $n_2$ denote the number of resolved and unresolved sources that generated the data, respectively.
For the resolvable population, we introduce the frequency-rescaled amplitude
\begin{equation}
    \tilde{A}=A(f/f_{\rm min})^{-2/3}.
\end{equation}
We expect this quantity to be equal to $A_0$ introduced in Eq.~\eqref{eq:def_strain}, within measurement errors, for all resolved sources. However, the fact that this value should be the same across all resolved sources is not enforced in the mock Global Fit. We therefore accommodate this by modelling the distribution of $\tilde{A}$ as a Gaussian with mean $A_P$ and standard deviation $\sigma_A$. Finally, we assume that the frequency is distributed according to a power-law.  Thus,
the resolvable population follows
\begin{equation}
    p_1(f,\tilde{A}|\Lambda)=\frac{{\rm PL}(f|\gamma_P,f_{\rm min},f_{\rm max})\alpha(f,f_{{\rm res},P},s_{{\rm res},P})}{p_{\rm res}(\gamma_P,f_{{\rm res},P},s_{{\rm res},P}) } \mathcal{N}(\tilde{A}|A_P,\sigma_{A}) ,
\end{equation}
where ${\rm PL}(f|\gamma_P,f_{\rm min},f_{\rm max})$ denotes a power-law distribution between $f_{\rm min}$ and $f_{\rm max}$ with exponent $\gamma_P$. We introduced the resolvability probability:
\begin{equation}
    p_{\rm res}(\gamma,f_{{\rm res}},s_{{\rm res}})=\int_{f_{\rm min}}^{f_{\rm max}} {\rm PL}(f|\gamma,f_{\rm min},f_{\rm max})\alpha(f,f_{{\rm res}},s_{{\rm res}}) \ {\rm d}f.
\end{equation}
The rate of resolvable events is
\begin{equation}
    N_1= p_{\rm res}(\gamma_P,f_{{\rm res},P},s_{{\rm res},P}) N.
\end{equation}
 For the background parameters, we have
\begin{equation}
    p(\Sigma|\Lambda)=p(A_{B}|\gamma_B,f_{{\rm res},B},s_{{\rm res},B},\Lambda)\mathcal{N}(\gamma_B|\gamma_P,\sigma_{\gamma}) \mathcal{N}(f_{{\rm res},B}|f_{{\rm res},P},\sigma_{f}) \mathcal{N}(s_{{\rm res},B}|s_{{\rm res},P},\sigma_{s}),
\end{equation}
where
\begin{align}
    p(A_{B}|\gamma_B,f_{{\rm res},B},s_{{\rm res},B},\Lambda)&=\sum_{n_2=0}^{+\infty} p(A_{B}|n_2,\gamma_B,f_{{\rm res},B},s_{{\rm res},B},\Lambda) p(n_2|\Lambda)  \nonumber \\
& = \sum_{n=0}^{+\infty} p \left( A_{B}\Bigg|{n}{(1-p_{\rm res}(\gamma_B,f_{{\rm res},B},s_{{\rm res},B}))},\gamma_B,f_{{\rm res},B},s_{{\rm res},B},\Lambda \right ) p(n|N)  \nonumber \\
& = p \left ( n=\frac{A_B^2(f_{\rm max}^{\gamma_B+1}-f_{\rm min}^{\gamma_B+1})f_{\rm min}^{4/3}}{(1-p_{\rm res}(\gamma_B,f_{{\rm res},B},s_{{\rm res},B}))(\gamma_B+1)(A_P^2+\sigma_A^2)\delta f} \Bigg | N \right ), \label{eq:p_af}
\end{align}
where we  used successively $n_2=n(1-p_{\rm res}(\gamma_B,f_{{\rm res},B},s_{{\rm res},B})) $ and
\begin{equation}
  p \left( A_{B}\Bigg|n_2,\gamma_B,f_{{\rm res},B},s_{{\rm res},B},\Lambda \right ) =\delta \left ( A_B-\sqrt{\frac{n_2(\gamma_B+1)(A_P^2+\sigma_A^2)\delta f}{(f_{{\rm max}}^{\gamma_B+1}-f_{{\rm min}}^{\gamma_B+1})f_{\rm min}^{4/3}} }\ \right )  .
\end{equation}
Since the Poisson likelihood is defined for integer numbers only, we take the integer part of the term inside the parenthesis when evaluating Eq.~\eqref{eq:p_af}. In this step, we use the background parameters measured in the mock Global Fit, $(\gamma_B,f_{{\rm res},B},s_{{\rm res},B})$, and not the population ones $(\gamma_B,f_{{\rm res},P},s_{{\rm res},P})$. In practice, each sample of these parameters is translated into a sample of the number of unresolvable events, and then into a sample of the total number of events, which is assumed to follow a Poisson likelihood.

Thus, the complete set of population hyperparameters is
\begin{equation}
  \Lambda=(\gamma_P,\sigma_{\gamma},A_P,\sigma_{A},N,f_{{\rm res},P},\sigma_{f},s_{{\rm res},P},\sigma_s) .
\end{equation}

\begin{figure}
    \centering
    \subfigure[Noise reconstruction]{
    \includegraphics[width=0.99\columnwidth]{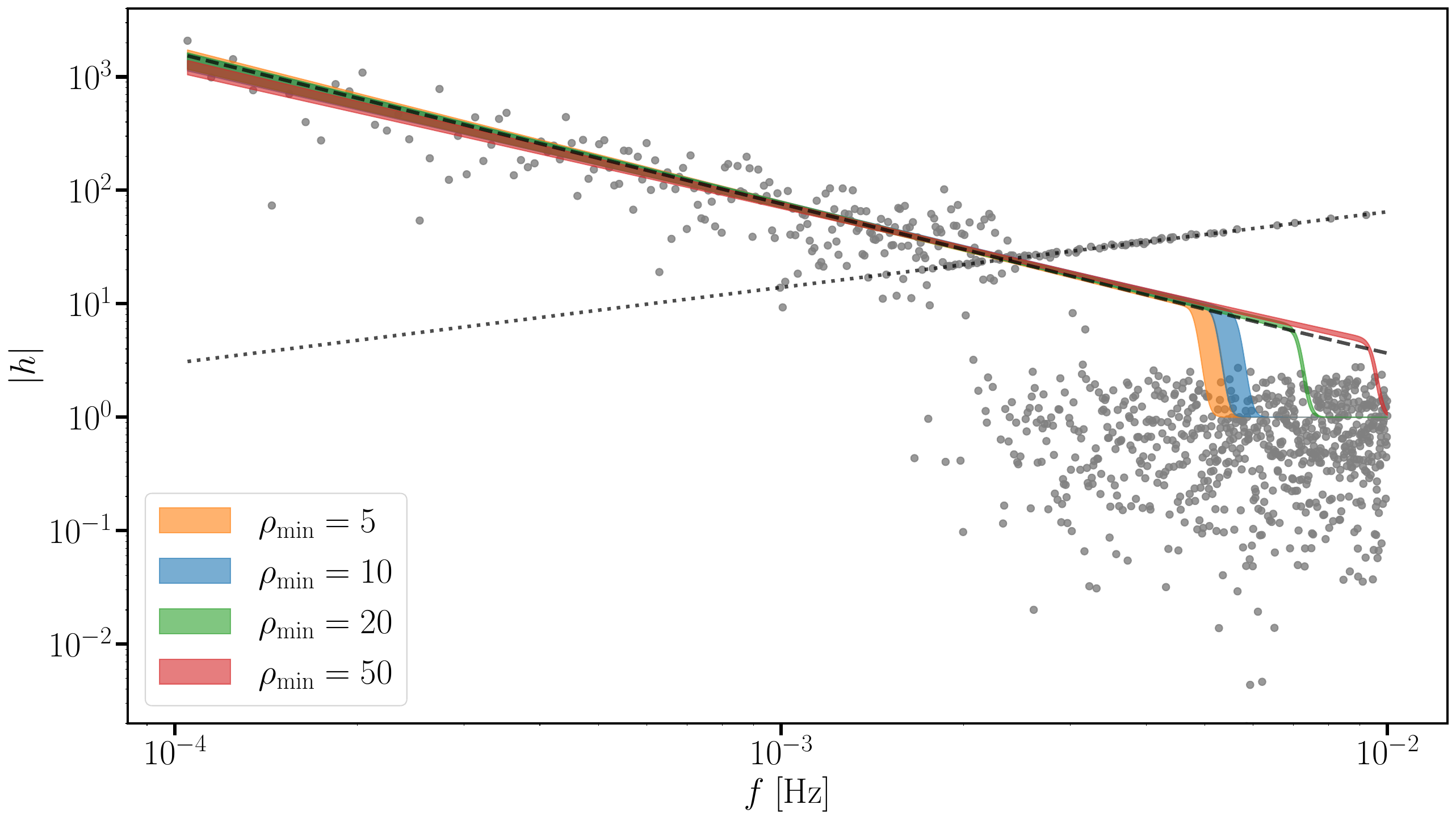}
    \label{fig:comp_ppd_noise}
    }
    \\

    \subfigure[Distribution of resolved sources]{
    \includegraphics[width=0.6\columnwidth]{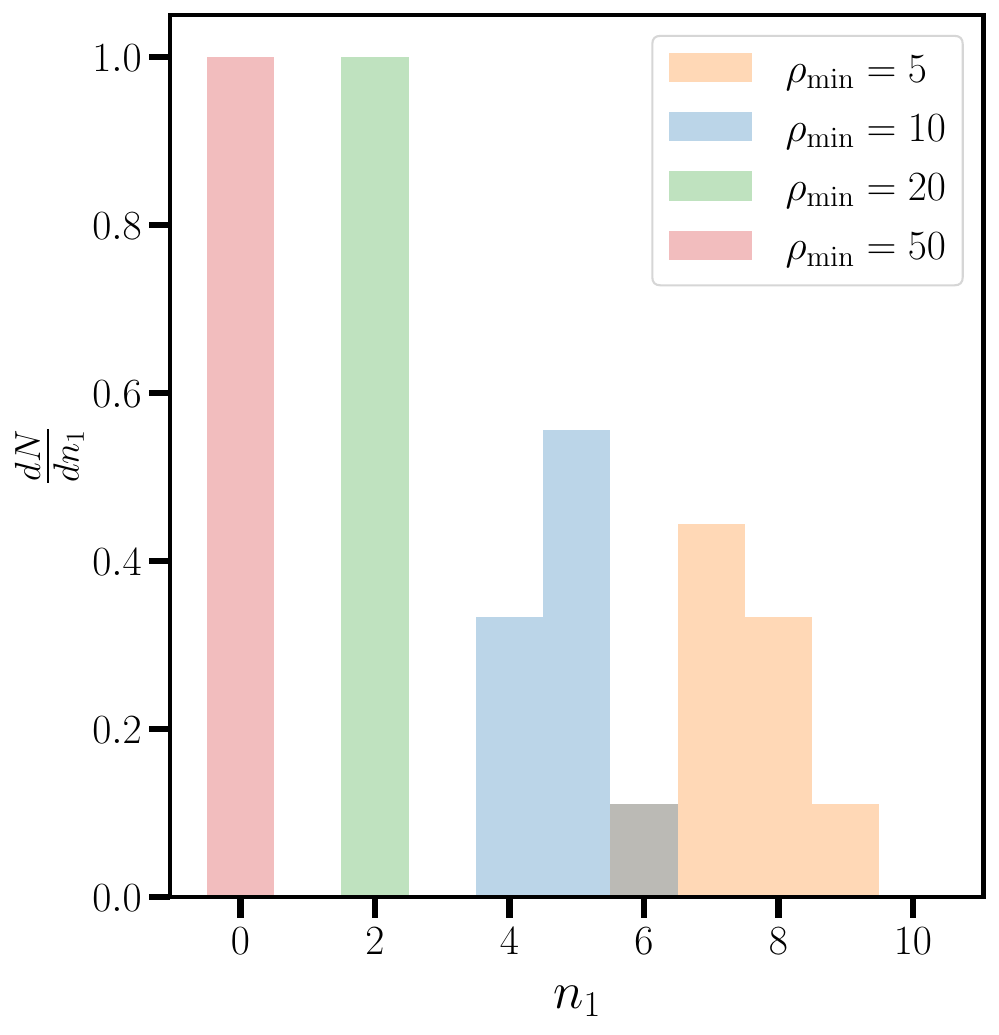}
    \label{fig:dist_nsource}
    }
    \caption{The upper panel shows the reconstruction of the total noise ($90\%$ confidence interval) and the lower panel the corresponding distribution on the number of resolved sources for different choices of the lower threshold on the SNR. Grey points show the data in all the bins, the black dashed line shows the theoretical expectation for the noise, given in Eq~\eqref{eq:sigma_foreground_th}, and the dotted one the amplitude of a single source. Increasing $\rho_{\rm min}$ decreases the number of resolved sources and pushes the background contribution to extend to higher frequency. }
\end{figure}

The original prior is converted into a prior on \mbox{$[\{f,\varphi,\tilde{A}\}_{n_1},(A_B,\gamma_B,f_{{\rm res},B},s_{{\rm res},B})]$} through the Jacobian transformation:
 \begin{equation}
     \pi_{\rm PE}(\{f,\varphi,\tilde{A})\}_{n_1},(A_B,\gamma_B,f_{{\rm res},B},s_{{\rm res},B} )) = \pi_{\rm PE} (\{f,\varphi,\rho\}_{n_1},(A_B,\gamma_B,f_{{\rm res},B},s_{{\rm res},B}) )\prod_{i=1}^{n_1} \frac{(f_i/f_{\rm min})^{2/3}} {\sigma^2_B(f_i)+\sigma_n^2}.
 \end{equation}
 For the population analysis, we assume a flat distribution for $\varphi$ between $-\pi$ and $\pi$,consistent with the mock Global Fit. Therefore, the $\varphi$ term cancels out in Eq.~\eqref{eq:lisa_pop_rjmcmc}, and $\varphi$ can be discarded from the population inference.

\subsection{Results}
We consider four different values of the minimum SNR of resolved sources: $\rho_{\rm min}=5,10,20,50$. We show in Fig.~\ref{fig:comp_ppd_noise} the data and the reconstruction of the total noise ($90\%$ confidence interval) for the different choices of $\rho_{\rm min}$, and in Fig.~\ref{fig:dist_nsource} the distribution of the number of resolved sources in each case. Bins with a single source are those aligning with the single-source amplitude shown with the black dotted line. At low frequencies, the reconstructed noise follows the theoretical prediction shown with the black dashed line, except in the $\rho_{\rm min}=50$ case. As we increase the SNR threshold, fewer sources are allowed to be resolved, and the background model has to compensate by extending to higher frequency. As the bins with single sources lie well above the background at high frequency, to accommodate this the recovered background slope flattens, which means it no longer matches the true distribution at low frequencies.

\begin{figure}
    \centering
    \includegraphics[width=0.6\columnwidth]{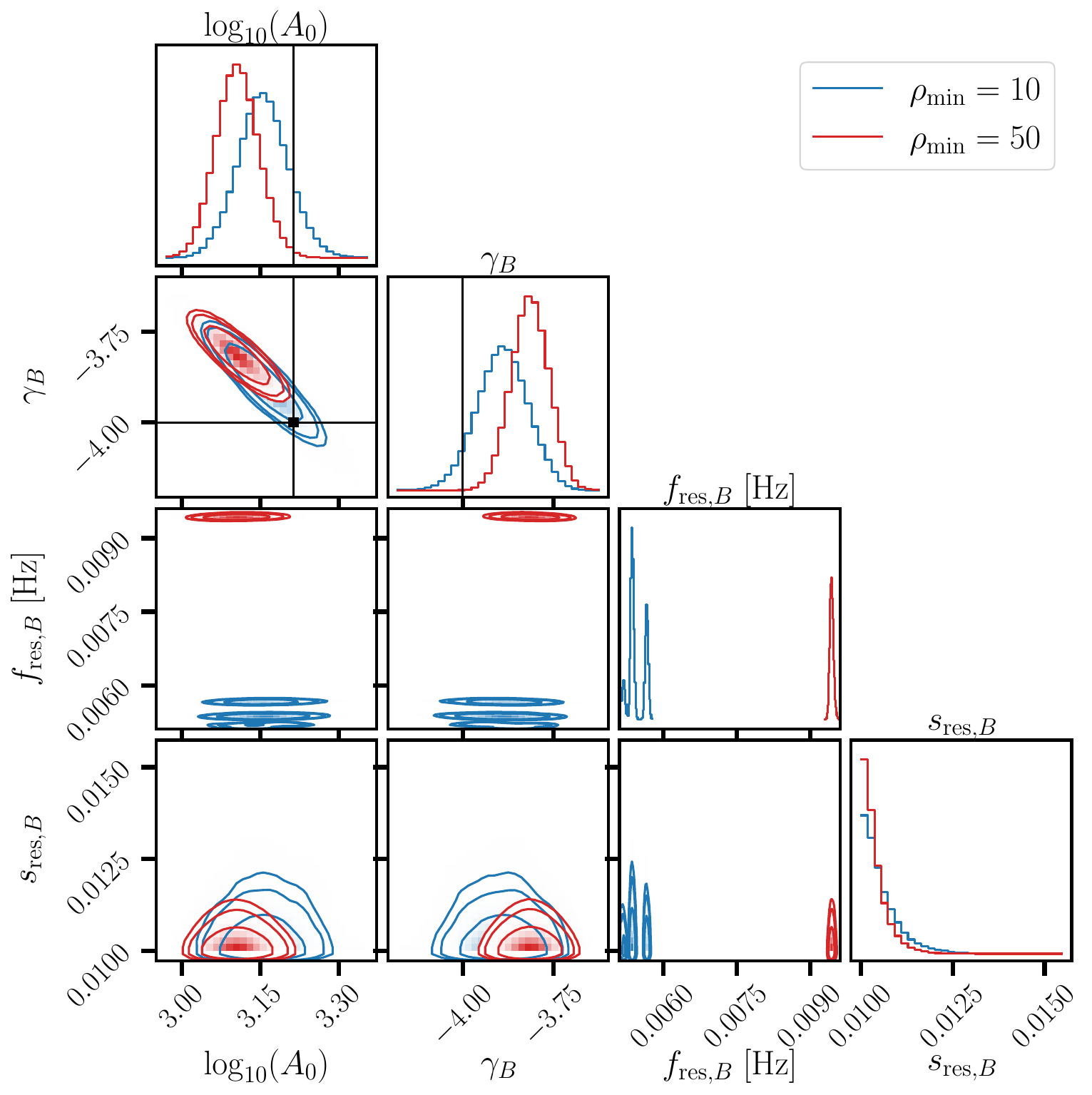}
    \caption{Posterior distribution of the noise parameters in the $\rho_{\rm min}=10$ (blue) and $50$ (red) cases. $A_0$ is the amplitude of the background in the first bin. We did not normalise the distributions of $f_{{\rm res},F}$ to the same binning scheme given their disjoint support in order to make the fine features more apparent.}
    \label{fig:comp_post_noise_pars}
\end{figure}

In the $\rho_{\rm min}=50$ case, no sources are resolved. Note that the SNR of the last bin containing a source is above 50 in the remaining cases. However, for $\rho_{\rm min}=50$ and as described above, the best fit is obtained by tilting the noise spectrum slightly upward at higher frequencies to compensate for not resolving the previous sources. As a consequence, for $\rho_{\rm min}=50$ the SNR of the last source ends up below 50, and it is therefore not resolved. We further illustrate this in Fig.~\ref{fig:comp_post_noise_pars}, where we compare the posterior on the noise parameters in the $\rho_{\rm min}=10$ and $\rho_{\rm min}=50$ cases. In the latter, the noise curve is estimated to be shallower (larger spectral index) with a lower amplitude in the first bin $A_0$, resulting in the tilt observed in Fig.~\ref{fig:comp_ppd_noise}. Moreover, in the $\rho_{\rm min}=10$ case, we observe that the $f_{{\rm res},B}$ distribution is multimodal, corresponding to the different numbers of resolved sources (see Fig.~\ref{fig:dist_nsource}). The more sources we resolve, the earlier the break in the noise curve. We find the distribution of $s_{{\rm res},B}$ to be peaked at 0 in all cases, favouring a sharp transition between the Gaussian and no-Gaussian regimes for our toy model.

\begin{figure}\label{eq:pres_emp}
    \centering
    \subfigure[]{
    \includegraphics[width=0.47\columnwidth]{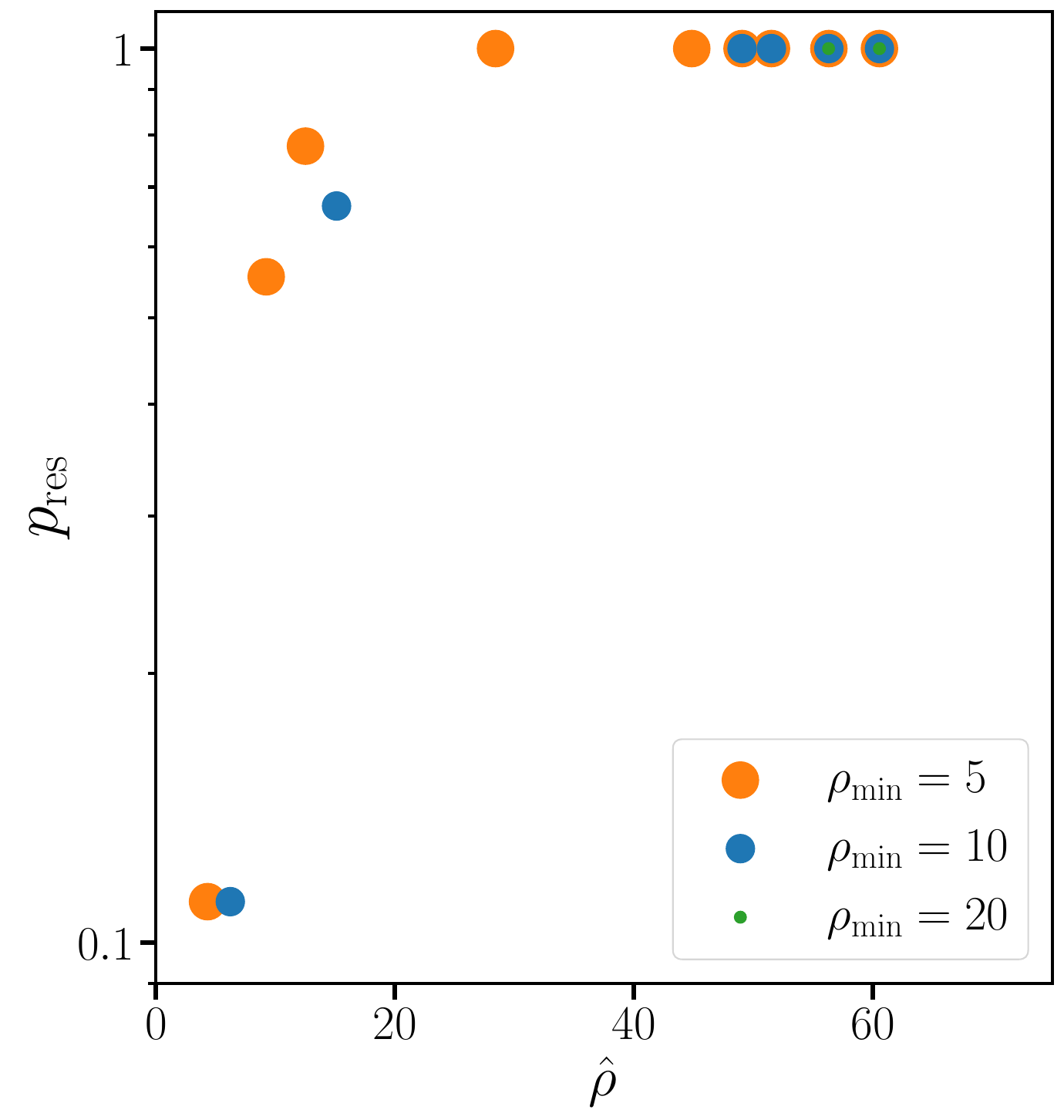}
    \label{fig:comp_post}
    }
    \subfigure[]{
    \includegraphics[width=0.47\columnwidth]{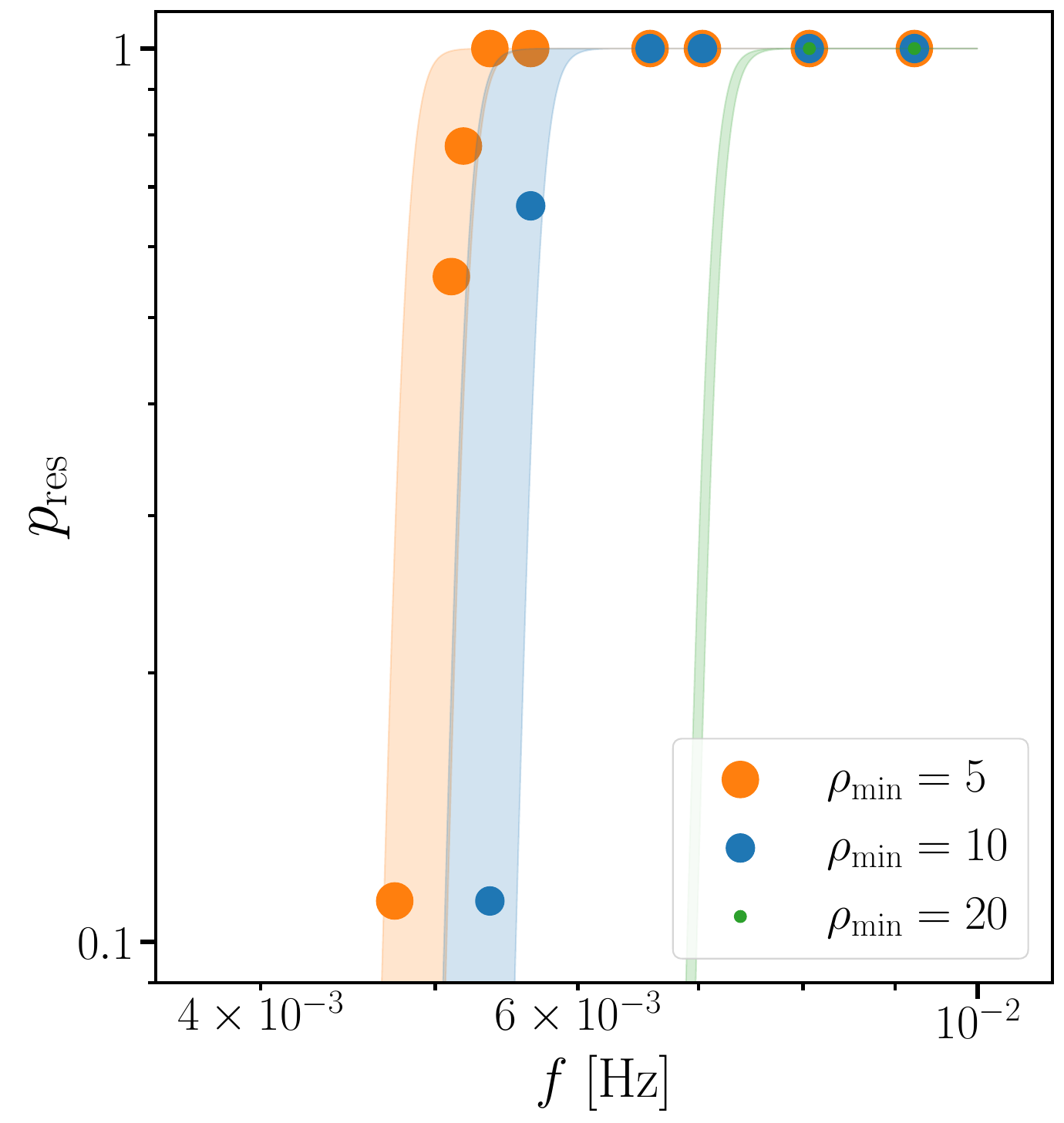}
    \label{fig:comp_cauchy}
    }
    \caption{Probability of resolvability of a source as a function their estimated SNR (left) and frequency (right). In the latter case we also show in coloured bands the $90\%$ confidence interval on $\alpha(f,f_{{\rm res}},s_{{\rm res}})$ defined in Eq.~\eqref{eq:res_func}.}
\end{figure}

Next, we show in Fig.~\ref{eq:pres_emp} the resolvability probability of a source as a function of its estimated SNR (left panel) and its frequency (right panel). The resolvability probability is estimated as the fraction of samples from our mock Global Fit in which that source is resolved. The estimated SNR $\hat{\rho}$ is computed by taking the ratio of the amplitude of the data in a given bin to the mean of the reconstructed noise in that bin. In both panels, we observe a clear monotonic trend.
On the right panel, we also compare the empirically estimated resolvability probability to the one computed from the resolvability function given in Eq.~\eqref{eq:res_func}. We find that they agree reasonably well. This is to be expected, since $\alpha$ defines the resolvability probability in our model. If the overall model fit was not good, for example because the background was far from Gaussian, then we might see inconsistencies, so we can consider this as a test of the goodness of fit of our model.

\begin{figure}
    \centering
    \includegraphics[width=0.98\columnwidth]{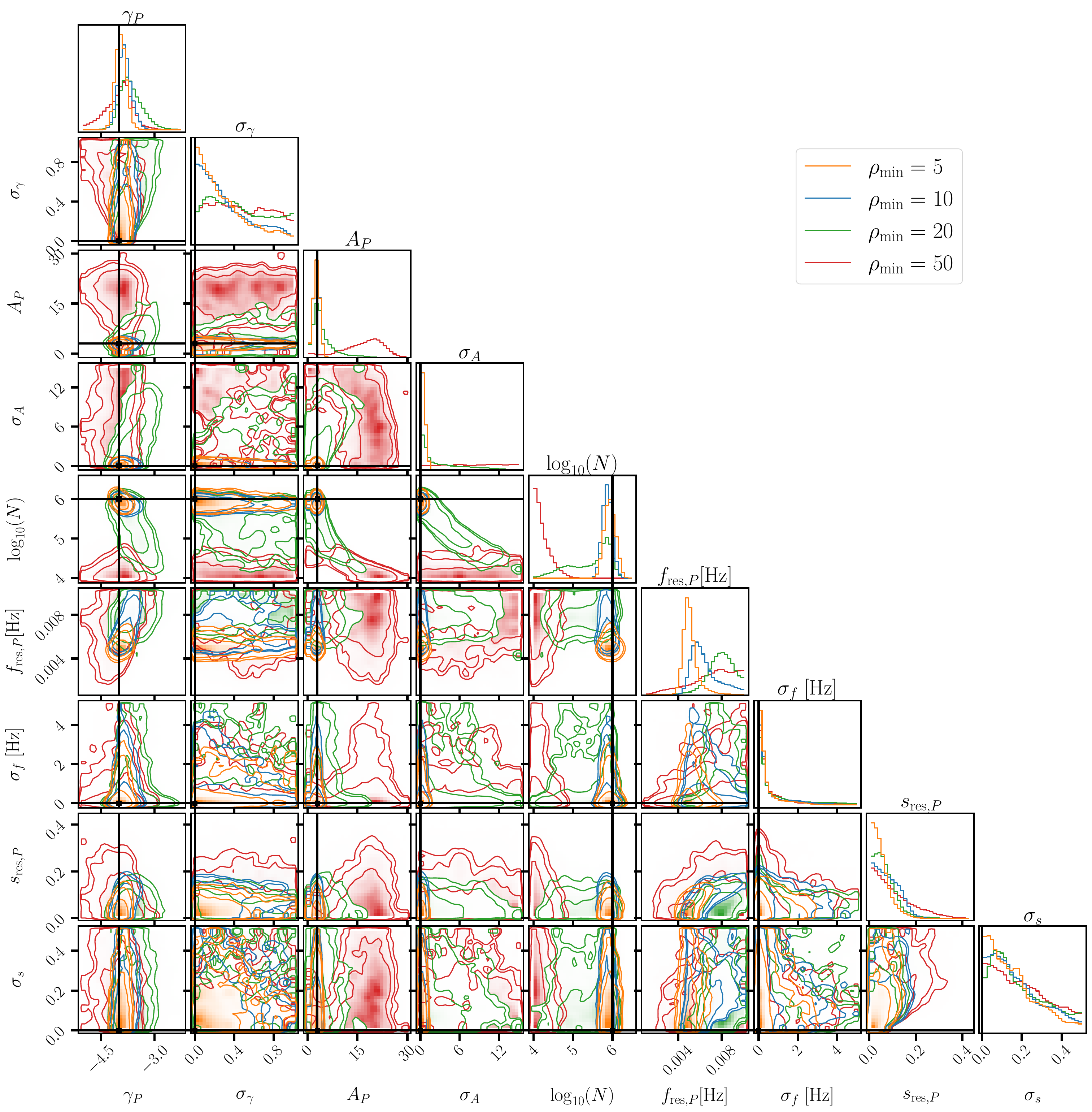}
    \caption{Posterior on the population hyperparameters for the different choices of $\rho_{\rm min}$ considered in this work. We show with a black line the true value of hyperparameters. For $f_{{\rm res},P}$ and $s_{{\rm res},P}$ there is no true value as they are fitted to optimise our description of the data as a stochastic and a resolvable component.}
    \label{fig:comp_pop_posterior}
\end{figure}

In the $\rho_{\rm min}=50$ case, the inference becomes biased, as already suggested by the posterior on the noise parameters in Fig.~\ref{fig:comp_post_noise_pars}. The posterior for $\gamma_P$ reflects two competing effects. On the one hand, the measurement of $\gamma_B$ favours shallower slopes. On the other hand, because no events are resolved, the Poisson term governing the number of resolved sources in Eq.~\eqref{eq:hyper_logl_1} pulls the inference toward a steeper slope and a smaller total number of events. The resulting posterior for $\gamma_P$ represents a compromise between these two influences and remains statistically compatible with the true value, but the inferred number of sources is heavily underestimated.
In this regime, the single-source amplitude is fully degenerate with the total number of sources: to reproduce the measured background amplitude, and in the absence of resolved events that would break the degeneracy, it is pushed toward larger values. Its standard deviation remains unconstrained, leading to a flat posterior. Because no information is available from the resolvable population, and thus no tension can arise between the resolved population and the background, he parameters $\sigma_f$ and $\sigma_s$ peak at zero. Finally, $\sigma_{\gamma}$ peaks slightly away from zero, consistent with the fact that the inferred $\gamma_P$ is pulled toward lower values relative to $\gamma_B$.

This latter result arises because adopting a minimum SNR threshold of 50 for resolved sources enforces a separation between the stochastic background and resolved sources that provides a poor description of the data. In this regime, the Gaussian-background approximation is extended well beyond where it is valid, and relying on such a mischaracterisation ultimately biases the population inference. Conversely, adopting a very low SNR threshold would increase the computational burden, as a larger number of sources would need to be treated as resolved. It could also introduce additional systematics, such as fitting noise artefacts as individual sources and increasing confusion between overlapping signals, potentially leading to multiple sources being incorrectly fitted as a single one. This, in turn, would also result in a poor description of the underlying population.

For the toy Global Fit we considered, choosing an SNR threshold between 5 and 20 allows us to describe the data accurately without making the fit excessively complicated (as would happen if the threshold were lowered) or too incorrect (as would happen if it were increased). We stress that these specific values apply only to our simplified setup and should not be directly translated into an assessment of the real LISA Global Fit. Nonetheless, through this toy model we have validated that the approach outlined for processing a Global Fit output, encapsulated in Eq.~\eqref{eq:hyper_logl_1}, can be used to estimate the parameters of the underlying population.

\section{Conclusions}
Given the tantalising scientific prospects of LISA for our understanding of astrophysical sources---particularly GBs, EMRIs, and massive black holes---being able to perform population analyses directly from the output of the Global Fit is essential. While current LVK analyses rely exclusively on information from individually resolved, non-overlapping events, and PTA constraints come primarily from the stochastic background, LISA will operate in an intermediate regime in which both individually resolvable sources and an overlapping background must be fitted jointly. As a result, correlations between these two components, naturally encoded in the Global Fit, need to be accounted for when reconstructing the underlying population.

In this work, we introduced an approach that coherently incorporates the correlation between the resolvable population and the background, as well as their joint relation to the astrophysical population, without requiring a post-processed catalogue of sources extracted from the Global Fit. In the Appendix, we also show how this general framework can approach the current LVK formalism, and discuss differences with the formalism proposed in~\cite{Callister:2020arv} to combine background and individual events. We outline how this formalism should evolve as the number of detected events and the overlap between them grow, particularly in anticipation of third-generation detectors such as the Einstein Telescope or Cosmic Explorer.

We applied our approach to a mock Global Fit as a proof of concept to illustrate how it can be implemented in practice.
In addition to a model for the population of sources, the key ingredients of our method are:
\begin{enumerate}
\item a resolvability function $\alpha(\theta, \Lambda)$;
\item a mapping between the properties of the stochastic background and those of the astrophysical population, $p(\Sigma \mid \Lambda)$.
\end{enumerate}
The first ingredient may be calibrated using a phenomenological fit, as we have done here, but studies of how different choices impact realistic LISA data will be required. For the second ingredient, the simplicity of our toy model allowed us to make an informed guess, but in realistic applications more sophisticated mappings will be needed, for instance, approaches based on deep learning~\cite{Giarda:2025ouf,DeSanti2026}.

Our approach uses the joint samples of resolved sources and of the background. Future studies will be needed to understand whether approaches based on catalogues of sources~\cite{Johnson:2025oyu} that do not track correlations between resolved sources and with the background allow for unbiased hierarchical inference. For completeness, let us mention that a fully deep-learning---based approach to inferring the population of GBs could provide an alternative that bypasses both the need for a Global Fit and a subsequent hierarchical Bayesian analysis~\cite{Srinivasan:2025etu}.

Another key aspect is the choice of SNR threshold for resolved events. Lower thresholds yield a more accurate representation of the data---ultimately better described as a superposition of individual sources---but the large expected number of LISA sources and their mutual confusion introduce severe degeneracies that render such thresholds impractical. Conversely, as the threshold increases, an increasingly large fraction of the dataset is forced to be fitted by the background model, including in frequency regimes where the data are not expected to be Gaussian at all, until the model eventually breaks and returns biased inferences. In a realistic case, no hard SNR cut is imposed but the choice of prior on the amplitude or the SNR of resolvable sources will similarly have an impact. Our results suggest that, for realistic astrophysical populations, hyperparameters influencing where the background ceases to be Gaussian will be particularly sensitive to all these choices. Allowing for flexibility in the resolvability function can partially compensate for these choices, but as we see here a poor choice of prior used in sampling can affect the final fit by leaving too much non-Gaussianity in the background component. Thus, care must be taken when choosing priors in the analysis and all of these choices must be propagated self-consistently throughout the inference (e.g.~when computing  $p(\Sigma \mid \Lambda)$).

An alternative to the approach discussed here would be to introduce a parametrised resolvability function directly within the Global Fit. This would let the prior on the parameters of resolved sources be adjusted to better represent data in terms of a Gaussian component and resolved sources, rather than using a fixed prior or fixed SNR threshold. Moreover, population inference could in principle be carried out directly within the Global Fit. In practice, this would amount to not performing the integration over $n_1$, $
\{\theta\}^{1}_{n_1}$, and $\Sigma$ in Eq.~\ref{eq:hyper_logl_1}. Instead, the population parameters would be inferred jointly with the parameters describing the individually resolved sources and the instrumental noise within a single Global Fit analysis. This approach is conceptually similar to what has recently been proposed in the LVK context, where only resolved sources are considered~\cite{Mancarella:2025uat}. More recently, Ref.~\cite{Criswell:2026xqk} presented an analysis that can be viewed as a particular implementation of this idea, in which the resolvability function is estimated empirically by drawing population realisations on the fly during the Global Fit. Performing a first population inference in the Global Fit using parametrised priors would also improve the efficiency of the Monte Carlo estimators used to evaluate the population likelihood, see Eq.~\eqref{eq:lisa_pop_rjmcmc}, one of the bottlenecks of current hierarchical analyses on LVK data~\cite{Talbot:2023pex,Heinzel:2025ogf}.

Samples obtained using a particular parametrised population model can be reweighted to alternative population models in post-processing in various ways. A first approach is to use Eq.~\eqref{eq:lisa_pop_rjmcmc}, allowing the prior on individual-event parameters to vary rather than remain fixed, which is not a problem within our formalism. Another possibility is to use a remapping technique. The full Global Fit output could be remapped by assigning pairs of individual-parameter and hyperparameter samples from the initially fitted model to the target model, based on a suitable distance metric between the target and first-fitted models. More simply, one could marginalise over individual event parameters, as usually done in population studies, and remap only the population parameters, following methods recently developed in the LVK context~\cite{Fabbri:2025faf,Rinaldi:2025evs} to map flexible non-parametric reconstructions onto parametric or astrophysically motivated models. Determining which regime enables precise astrophysical inference, and developing the tools required for realistic LISA data, will be the subject of future work.

Finally, how our approach can be extended to include other sources, in particular massive black hole binaries and extreme-mass-ratio inspirals (EMRIs) and whether injection campaigns as used by the LVK~\cite{Essick:2025zed} will be needed will also be the subject of future work. In considering other source types, an additional complication arises due to the size of the signal parameter space. The formalism here relies on the fact that it is possible to obtain converged RJMCMC samples over the full model space. In practice, the EMRI parameter space is so large that this is unlikely to be possible~\cite{Gair:2004iv,Babak:2009ua}. Instead, the analysis of the data will be separated into a search phase where EMRIs are identified in the data, followed by a parameter inference phase where the posteriors are obtained while simultaneously adjusting the parameters of all sources in the Global Fit. In such a set up, the separation between resolved sources and background is not purely driven by the data, but also by the effectiveness of the search pipeline. This inference model can also be accommodated in our pipeline, with the resolvability function now determined by the probability that the search identifies a source with given parameters. This can be determined on the fly as the search proceeds, or by an injection campaign, and then kept fixed or allowed to adjust slightly during the refinement phase. While technically there is a deterministic mapping between the observed data and the number of EMRIs that will be found and then fitted, the size of the EMRI parameter space is large enough that it should be possible to treat this probabilistically (as in the framework here), with the implied average over data realisations replaced by an average over the parameter space. A full exploration of the EMRI population fitting problem, and introduction of any necessary refinements, should be the subject of future studies, but the formalism we have described here should provide a valid framework also for that work.

\acknowledgments

We are thankful to G. Faggioli for his contribution in the early stages of this work and to S. Babak, R. Buscicchio, A. Criswell and N. Karnesis for fruitful discussions. A.T. is supported by MUR Young Researchers Grant No. SOE2024-0000125, ERC Starting Grant No.~945155--GWmining, Cariplo Foundation Grant No.~2021-0555, MUR PRIN Grant No.~2022-Z9X4XS, Italian-French University (UIF/UFI) Grant No.~2025-C3-386, MUR Grant ``Progetto Dipartimenti di Eccellenza 2023-2027'' (BiCoQ), and the ICSC National Research Centre funded by NextGenerationEU.
\clearpage
{\centering\textbf{\huge Appendix}
\vspace{0.1in}}
\appendix

\section{LVK limit}\label{app:lvk_limit}

The underlying assumption of the population analysis as used in the LVK is that the data has been divided into segments, and that a very large number of segments, $K\gg 1$, have been observed. {We apply our treatment of probabilistic inclusion of sources to each segment, first considering only resolvable-like sources. Denoting the expected number of events in the Universe that would occur during the observation time by $N$, the data in segment $i$ by $d_{c,i}$, and introducing flags $\{f_i\}$ that indicate the segments that contain ($f_i=1$) or do not contain ($f_i =0$) events, the joint likelihood can be written as
\begin{equation}\label{eq:lvk_prob}
p(d,\{\theta\}_K|\Lambda) \propto \prod_{i=1}^K \left[ f_i  p(d_{c,i} | \theta_i) p(\theta_i|\Lambda)+ (1-f_i) p_n(d_{c,i})\right]
\end{equation}
where $p_n(d_c)$ is the likelihood of the noise process. The prior on the flags is a binomial distribution with $p=N/K$. Marginalising over this prior gives instead
\begin{equation}
p(d,\{\theta\}_K|\Lambda) \propto \prod_{i=1}^K \left[ \frac{N}{K} p(d_{c,i} | \theta_i)p(\theta_i|\Lambda) + \left(1-\frac{N}{K}\right) p_n(d_{c,i})\right].
\end{equation}

The form of the model used in our formalism is inherently probabilistic. All data segments are sometimes used for inference of resolved sources and sometimes not. The standard LVK approach on the other hand discards data segments that are thought to contain no events based on some criterion, i.e.,~ data is selected.
Working with Eq.~(\ref{eq:lvk_prob}) and obtaining the posterior on a given flag, $f_i$, we can see that the ratio of the probability that $f_i=1$ to the probability that $f_i=0$ is just the Bayes factor for the resolved source model versus instrumental noise model in a given segment. This Bayes factor depends on the data (i.e., on the parameters that generated the event and on the noise realisation, see~\cite{Essick:2023upv}), but also on the population parameters $\Lambda$. The LVK approach uses a fixed population model for computing the detection statistic, but in principle selection on data should be carried out jointly with the hierarchical inference. Assuming the threshold is chosen high enough that the population model has relatively little impact on which segments are selected, then we can separate data segments into those that are above this threshold and those that are below the threshold. We additionally assume that in the segments above threshold $p(d_{c,i} | \theta_i)  \gg p_n(d_{c,i})$. Marginalising over the data and parameters in the other segments, and assuming that the threshold has been chosen so that the noise generating likelihood has no support in the ``detected'' data region, gives us
\begin{align}
p(\{d_c\}_{n_{\rm det}}|\Lambda)& \propto \left[\prod_{i=1}^{n_{\rm det}}  \frac{N}{K} \int p(d_{c,i} | \theta) p(\theta|\Lambda) {\rm d} \theta \right] \left[ \frac{N}{K} p_{\rm ndet}(\Lambda) + \left(1-\frac{N}{K}\right) \right]^{K-n_{\rm det}} \nonumber \\
& \propto \left[\prod_{i=1}^{n_{\rm det}}  \frac{N}{K} \int p(d_{c,i} | \theta) p(\theta_i|\Lambda) {\rm d}\theta \right] e^{N p_{\rm ndet}(\Lambda) - N},
\end{align}
where the last line follows from taking the limit as $K\rightarrow \infty$, and we introduced
\begin{align}
    p_{\rm ndet}(\Lambda) & = \int_{d_c \in {\rm ndet}} p(d_c|\theta)p(\theta|\Lambda)\  {\rm d}d_c {\rm d}\theta  \\
    & =1 -\xi(\Lambda) \nonumber
\end{align}
Finally, we get
\begin{equation}
p(\{d_c\}_{n_{\rm det}}|\Lambda) \propto e^{- N\xi(\Lambda)}  \prod_{i=1}^{n_{\rm det}} \int p(d_{c,i} | \theta) N(\theta_i|\Lambda) {\rm d}\theta  \\
\end{equation}
We see that we recover the usual expression as expected.

We now include a second process that generates background-like events, with also at most one such event per segment. We discuss afterwards the more realistic case of having multiple background-contributing events per segment. We allow a priori for having both resolved and background events in the same segment, but the cross-terms this introduces will disappear when we take the limit $K \rightarrow \infty$.
Following our previous notation, we note the population priors on resolved and background events $p_1(\theta|\Lambda)$ and $p_2(\theta|\Lambda)$, respectively. They are related to the global population prior through the resolvability function $\alpha(\theta,\Lambda)$, as in Eqs.~\eqref{eq:p1def} and \eqref{eq:p2def}. We introduce a second flag, $b_i$, which is $1$ for segments that include unresolved events, and $0$ for segments that do not. The first flag, $f_i$, will indicate the presence or absence of resolved events. The joint likelihood now reads
\begin{align}
p(d|\Lambda) \propto \prod_{i=1}^K \left[ f_i p_1(d_{c,i} | \Lambda) + (1-f_i) b_i p_2(d_{c,i} | \Lambda) + (1-f_i)(1- b_i) p_n(d_{c,i})\right],
\label{eq:withflags}
\end{align}
where
\begin{equation}
p_{1,2}(d_{c} | \Lambda) = \int p_{1,2}(d_{c} | \theta_i,\Lambda) p_{1,2}(\theta|\Lambda).
\end{equation}
Note, that in a typical situation the likelihood does not differ between resolved and unresolved sources, so $p_{1}(d_{c,i} | \theta_i)\equiv p_{2}(d_{c,i} | \theta_i)$. In the above we allow for the likelihoods to be different because this allows for greater generality. We include a $\Lambda$ dependency in the likelihoods, as, in all generality, the total noise model used there depends on $\Lambda$ through the background.

The priors on the two sets of flags are binomial distributions with $p=N_1/K$ and $p=N_2/K$ respectively. Marginalising over both the flags and parameters of individual sources we obtain
\begin{align}
p(d|\Lambda) \propto \prod_{i=1}^K \left[ \frac{N_1}{K} p_1(d_{c,i} | \Lambda)  + \left(1-\frac{N_1}{K}\right) \frac{N_2}{K} p_2(d_{c,i} | \Lambda) + \left(1-\frac{N_1}{K}\right) \left(1-\frac{N_2}{K}\right)p_n(d_{c,i})\right].
\label{eq:segmentlikeLowrate}
\end{align}

Expanding this product, we see that each term includes a set of $0 \leq N_{\rm obs} \leq K$ resolved source terms, coming from the first term in the bracket, combined with another term of the form
\begin{equation}
\left(1-\frac{N_1}{K}\right)^{K-n_{\rm det}} \prod_{j=1}^{K-n_{\rm det}} \left[  \frac{N_2}{K} p_2(d_{c,j} | \Lambda) +  \left(1-\frac{N_2}{K}\right)p_n(d_{c,j})\right].
\end{equation}
The first part gives $\exp(-N_1)$ in the limit $K \rightarrow \infty$. The second part is the part we choose to approximate as a stochastic process. Replacing this by a Gaussian this becomes the likelihood of the data given the Gaussian variance.

The ratio of the probability that $f_i=1$ to the probability that $f_i=0$ is now the Bayes factor for the resolved source model versus the background plus instrumental noise model. We threshold as before on this Bayes factor, and separate data segments into those above threshold and those below threshold. If we assume additionally that this threshold is high enough that the resolved source term dominates for those data segments, these terms contribute
\begin{equation}
\prod_{i=1}^{n_{\rm det}} \frac{N_1}{K} p_1(d_{c,i}|\Lambda).
\end{equation}
This can be simplified by defining
\begin{align}
    \tilde{p}_{1,2}( d_c|\Lambda) &=\frac{N_{1,2}}{N} p_{1,2} (d_c|\Lambda),
\end{align}
which is just the numerator part of Eqs.~\eqref{eq:p1def} and \eqref{eq:p2def}. The resolved source term becomes
\begin{equation}
\prod_{i=1}^{n_{\rm det}} \frac{N}{K} \tilde{p}_1(d_{c,i}|\Lambda).
\end{equation}
To handle the other terms we can treat these as they are treated in the LVK case, which is to ignore them by marginalising over the data in each segment. This marginalisation gives, for each term,
\begin{equation}
\int_{d_c \in {\rm ndet} } \frac{N}{K} \left (\tilde{p}_1(d_c|\Lambda) + \tilde{p}_2( d_c|\Lambda) \right) + \left(1-\frac{N}{K}\right) p_n({d_c}) \ {\rm d}{d_c}
\end{equation}
where we have dropped terms that are quadratic in $N/K$. If we assume that the threshold is high enough that noise cannot generate data in the detection-region the integral of $p_n(d_c)$ over the non-detection region is equal to unity and this term becomes
\begin{equation}
    \frac{N}{K} \left (1-\xi(\Lambda) \right) + \left(1-\frac{N}{K} \right) = 1 - \frac{N}{K} \xi(\Lambda),
\end{equation}
which follows because the integral of $\tilde{p}_1(d_c|\Lambda) + \tilde{p}_2( d_c|\Lambda)$ yields the total fraction of non detected events $1-\xi(\Lambda)$).
There are $K-N_{\rm obs}$ of these terms, and so we have
\begin{equation}
\left(1 - \frac{N}{K} \xi(\Lambda) \right)^{K-n_{\rm det}} \rightarrow e^{-N\xi(\Lambda)} \qquad \mbox{as} \qquad K\rightarrow\infty.
\end{equation}
Putting the pieces together, we get
\begin{equation}
    p(\{d_c\}_{n_{\rm det}}|\Lambda) \propto e^{-N\xi(\Lambda)}  \prod_{i=1}^{n_{\rm det}}  \int p_1(d_{c,i}|\Lambda) N_1(\theta|\Lambda).
\end{equation}
We see that we recover the standard LVK expression, with the exception that the resolved sources contribute $p_1(\theta|\Lambda)$, which depends on $\alpha(\theta,\Lambda)$. However, with the assumption that the segments above threshold clearly contain resolved sources, alpha must be approximately equal to 1 for all the parameter values generating data in this region.

In the above we have assumed that segments contain not only at most one resolved source but also at most one background source.
This assumption is not required, however, and can be eliminated by replacing Eq.~(\ref{eq:segmentlikeLowrate}) with
\begin{equation}
\prod_{i=1}^K \left[ \frac{N_1}{K} \tilde{p}_1(d_{c,i} | \Lambda) + \left(1-\frac{N_1}{K}\right) q(d_{c,i}|\Lambda)\right]
\label{eq:segmentlikehighrate}
\end{equation}
where
\begin{align}
q(d_c|\Lambda) &= \sum_{m=0}^{\infty}\left(\frac{N_2}{K}\right)^m \frac{1}{m!} {\rm e}^{-N_2/K} p_m(d_c|\Lambda) \nonumber \\
p_m(d_c|\Lambda) &= \int p_n\left(d_c - \sum_{i=1}^{m}h(\theta_i)\right) \prod_{i=1}^m \frac{p(\theta_i|\Lambda)(1-\alpha(\theta_i,\Lambda))}{N_2/N} {\rm d}\theta_i , \label{eq:noise_like_lvk}
\end{align}
with $h(\theta)$ being the GW signal of a source.
Marginalising over non-detectable data as before, and again assuming that the probability that the background process generates data in the detected region is vanishingly small, so that the integral of $q(d_c|\Lambda)$ over that region is $1$, we obtain instead of Eq.~\eqref{eq:int_ndet}
\begin{equation}\label{eq:int_ndet}
    \frac{N}{K} \left(\frac{N_1}{N} - \xi_{1} (\Lambda)\right) + \left(1-\frac{N}{K}\right) = 1 - \frac{N}{K} \xi_{1} (\Lambda)
\end{equation}
where
\begin{equation}
    \xi_{1}(\Lambda) = \int_{d_c \in {\rm det}} p(d_c|\theta) p(\theta|\Lambda) \alpha(\theta,\Lambda) {\rm d} \ d_c.
\end{equation}
Under the assumptions we have made, we can assume that $\alpha \approx 1$ in the region of parameter space generating data in the detectable region and so $\xi_{1}(\Lambda) \approx \xi(\Lambda)$ and so we once again obtain the standard result.

So far we have been marginalising out the below-threshold data. However, this means we are throwing away all the information that they provide about the total noise generated by the background plus instrument, which we want to keep. If we retain these terms the contribution from the data segments that are not in the clearly resolved region defined by our Bayes factor threshold is
\begin{equation}
\prod_{i=1}^{K-n_{\rm det}} \left[ \frac{N}{K} \tilde{p}_1(d_{c,i} | \Lambda) + \left(1-\frac{N_1}{K}\right) q(d_{c,i}|\Lambda)\right].
\label{eq:bkgrd}
\end{equation}
To do things correctly we need to explicitly keep all these terms, but sampling all of them would be expensive. One approximation would be to assume that for all of these terms only the $q(d_c|\Lambda)$ part is important. This is not true, however, as the transition from resolved to background sources should not be abrupt. If we did do this the prefactors $(1-N_1/K)^{K-n_{\rm det}}$ would contribute a term $\exp(-N_1)$. However, this is not consistent since we are keeping some terms proportional to $N_1/K$ and not others. Alternatively, the resolved source term can be averaged over data, so that these segments do not have to be fully analysed with a resolved source model, but we retain the data dependence in the second term. To do this we write the above expression as
\begin{align}
    &\left(1-\frac{N_1}{K}\right)^{K-n_{\rm det}}\prod_{i=1}^{K-n_{\rm det}} q({d}_{c,i}|\Lambda) \prod_{j=1}^{K-n_{\rm det}}\left[1 + \frac{N}{(K-N_1)} \frac{\tilde{p}_1({d}_{c,i} | \Lambda)}{q({d}_{c,i}|\Lambda)} \right] \nonumber \\
    &\hspace{2cm} \equiv {\rm e}^{-N_1} p(\{{d}_c\}_{\rm rest}|\Lambda) \prod_{j=1}^{K-n_{\rm det}}\left[1 + \frac{N}{(K-N_1)} \frac{\tilde{p}_1({d}_{c,i} | \Lambda)}{q({d}_{c,i}|\Lambda)} \right] \label{eq:margwithnoise}
\end{align}
where in the last line we are defining the ``noise-likelihood'' term $p(\{d_c\}_{\rm rest}|\Lambda)$, which appeared as $\int p(d|\Sigma) p(\Sigma |\Lambda) \ {\rm d}\Sigma$ in Eq.~(\ref{eq:pop_res_stoch_std}) and we took the $K\rightarrow \infty$ limit for the first term. We now eliminate the final product term by replacing the term in square brackets with its expectation value. This can be justified by the central limit theorem --- the product of a large number of independent and identically distributed random variables becomes increasingly concentrated around its expectation value to the power of the number of observations as that number tends to infinity. The expectation value can be computed as
\begin{equation}
   1+ \frac{K}{K-N \xi_{1}(\Lambda)} \int_{d_c \in {\rm ndet}} \left( \frac{N}{K} \tilde{p}_1(d_{c} | \Lambda) + \left(1-\frac{N_1}{K}\right) q(d_{c}|\Lambda)\right) \frac{N}{K-N_1} \frac{\tilde{p}_1({d}_{c} | \Lambda)}{q({d}_{c}|\Lambda)} \,{\rm d}d_c
\end{equation}
The factor outside the integral is the normalisation that comes from the fact that these segments are restricted to be in the non-detected region and so the expectation is taken over the distribution of data within that region (it is one over the integral of the term in brackets over that region). We see that to $O((N/K)^2)$ the contribution is
\begin{equation}
    1 + \frac{N}{K} \int_{d_c \in {\rm ndet}} \tilde{p}_1({d}_{c} | \Lambda) \,{\rm d}d_c = 1 + \frac{N}{K} \left(\frac{N_1}{N} - \xi_{1}(\Lambda) \right).
\end{equation}
Replacing each square bracketed term in Eq.~(\ref{eq:margwithnoise}) with this expression and then taking the usual limit $K\rightarrow \infty$ we obtain
\begin{equation}
    {\rm e}^{-N_1} p(\{{d}_c\}_{\rm rest}|\Lambda) {\rm e}^{N_1 - N \xi_{1}(\Lambda)} = p(\{{d}_c\}_{\rm rest}|\Lambda) {\rm e}^{- N \xi_{1}(\Lambda)}.
\end{equation}
Putting the pieces together, we get
\begin{equation}
    p(d|\Lambda) =  p(\{{d}_c\}_{\rm rest}|\Lambda) {\rm e}^{- N \xi_{1}(\Lambda)} \prod_{i=1}^{n_{\rm det}}  \int p_1(d_{c,i}|\Lambda) N_1(\theta|\Lambda),
\end{equation}
which takes the same form as Eq.~(\ref{eq:pop_res_stoch_std}) once we make, as before, the identification that $\xi_{1}(\Lambda) \approx \xi(\Lambda)$.

In this way we have shown how to get results that look like the LVK result from our formalism in the appropriate limit. One important thing to note is that the ``noise likelihood'' term, $ p(\{{d}_c\}_{\rm rest}|\Lambda)$, in the above equations comes only from the unresolved population, see Eq.~\eqref{eq:noise_like_lvk}, and the detected term comes from the resolved one, while in Eq.~(\ref{eq:pop_res_stoch_std}), the whole population is used in both terms (since the split between resolved and unresolved sources was not explicitly done). In the current LVK setup $\alpha(\theta,\Lambda)\sim 1$ and it can be considered constant (though not exactly 1) over the range of parameter space contributing to the likelihood, so that $p_{1,2}(\theta|\Lambda)\simeq p(\theta|\Lambda)$, and we recover Eq.~\eqref{eq:noise_like_lvk}. However, this is valid only while the background is below the detection threshold. Once a background is measurable, this split between resolved and unresolved sources should be taken into account and the resolvability function should be estimated from data. As the resolvability threshold will be driven by the goodness-of-fit of the assumed background model to the unresolved population, we expect this to occur at a point where there are multiple resolved sources within each segment, since there will have to be multiple unresolved sources contributing to the background to make it appear stochastic. This is the main reason why the current LVK formalism will need to be modified in the future.

This derivation has illustrated where approximations are made in the LVK approach, in particular the fact that we assume that there are a very large number of data segments, and that resolved sources are rare so that no segment can contain more than one. For future detectors these assumptions will become increasingly poor and so the standard LVK analysis will no longer be useable. The formalism described here can be readily applied to that case, however. The below-threshold segments in the above expression could potentially be handled in other ways, for example using a Gaussian approximation and the central limit theorem. The contribution to the Gaussian variance would then come from both resolved and background events and hence depend on the total rate $N$. In any approach that does not do any marginalisation over unresolved segments the whole product over these segments (as in Eq.~\eqref{eq:margwithnoise}) can be thought of as $p(\{d_{c}\}_{\rm rest}|\Lambda)$, but in that case the $\exp(-N\xi(\Lambda))$ term that appears in Eq.~(\ref{eq:pop_res_stoch_std}) should not be there. This term arises because of the marginalisation over the data segments that are not being used for the resolved source analysis. Once we add the background term we are reintroducing that data and so it has not been marginalised out, therefore this term should no longer appear.

Finally, we note that the separation of segments unto those that definitely have resolved events and others that are ambiguous was only done to connect to the LVK limit, and because it reduces computational cost when events are rare. As rates increase, or signals become longer (as in the LISA case), this separation no longer makes sense as most of the data will be being used anyway and segments may contain more than one resolved event. Again, the formalism described here can be directly applied in any general context without problem. If we sample from the full model described here, including the flags (for example using reversible jump to move between the $f_1=0$ and $f_i=1$ models), then all the segments will include a resolved source a fraction of the time proportional to the Bayes factor. This can be thought of as a variant of the LVK-type analysis in which data segments are chosen probabilistically, rather than with a fixed threshold. Such an approach is in principle able to extract more information from the data, but at the price of significantly higher computational cost since parameter estimation has to be done in every segment that has non-trivial Bayes factor.

\bibliography{ref.bib}
\bibliographystyle{ieeetr}

\end{document}